\documentclass[a4paper,11pt]{article}
\usepackage{jheppub} 
\usepackage{lineno}

\usepackage[english]{babel}
\usepackage{graphicx} 
\usepackage{multicol}
\usepackage{float}
\usepackage{booktabs} 
\usepackage{amsthm, amsmath, amssymb, bm, ascmac, multicol, centernot, mathtools, autobreak, physics, cancel}
\usepackage{feynmp-auto}
\usepackage[normalem]{ulem} 
\usepackage{comment}
\usepackage[nameinlink,capitalize]{cleveref} 
\usepackage{tikz}
\usetikzlibrary{intersections,calc,arrows.meta}

\usepackage[most]{tcolorbox}
\newtcolorbox{qbox}[2][]{
  colback=white, colframe=black,
  fonttitle=\bfseries,
  title=Question #2,
  #1
}

\newcommand\Dcal{\mathcal{D}}
\newcommand\Lcal{\mathcal{L}}
\newcommand\Ocal{\mathcal{O}}
\newcommand\Mcal{\mathcal{M}}
\newcommand\im{\mathrm{i}}

\newcommand\p{\partial}
\newcommand\rd{{\rm d}}

\newcommand{\changed}[1]{#1}

\title{\boldmath
    Effective field theory for dissipative photons \\
    from higher-form symmetries
}

\author[a]{Genki Yoshimura,}
\emailAdd{gyoshimura@kern.phys.sci.osaka-u.ac.jp}

\author[a]{Yukinao Akamatsu}
\emailAdd{yukinao.a.phys@gmail.com}

\author[b]{and Yuji Hirono}
\emailAdd{hirono@iit.tsukuba.ac.jp}

\affiliation[a]{
    Department of Physics, The University of Osaka,\\
    1-1 Machikaneyama, Toyonaka, Osaka 650-0043, Japan
}

\affiliation[b]{
    Institute of Systems and Information Engineering, University of Tsukuba,\\
    1-1-1 Tennōdai, Tsukuba, Ibaraki 305-8573, Japan
}

\abstract{
Recent developments in generalized symmetries have provided new insights
into quantum field theories. Within this framework, photons can be understood
as Nambu--Goldstone modes associated with a spontaneously broken higher-form symmetry.
In this work, we develop an effective field theory that
builds on this symmetry structure to describe the real-time dynamics of
photons in insulating media at finite temperature.
Combining the Schwinger--Keldysh formalism with the generalized coset
construction, we formulate a symmetry-based effective action that
incorporates both conservative and dissipative effects.
The effective theory implements the dynamical Kubo--Martin--Schwinger symmetry,
ensuring consistency with the fluctuation--dissipation relation and Onsager's
reciprocal relations. Within this framework, we derive the entropy current associated with
dissipative photon dynamics and demonstrate the non-negativity of its
divergence, in accordance with the second law of thermodynamics.
We also clarify the symmetry origin of the gauge redundancy in the unbroken phase within the Schwinger--Keldysh framework, relating it to strong and weak realizations of higher-form symmetries.
Our results provide a model-independent effective description of photon dynamics in insulating media at finite temperature.
}

\begin{document}
\maketitle
\flushbottom

\section{Introduction}

Symmetry is a universal structural feature
of physical systems
and provides a powerful organizing principle, independent of microscopic details. It plays a central role in our understanding of a wide range of physical phenomena. A prominent example is the Landau paradigm, in which phases of matter are classified according to their symmetry properties~\cite{Landau:1937obd,LandauLifshitz5}.

Effective Field Theory (EFT) is a general low-energy framework that describes the long-distance dynamics of systems with a given symmetry structure~\cite{Weinberg:1996kr}. It proceeds in three main steps: identifying the relevant low-energy degrees of freedom, constructing the most general Lagrangian consistent with the symmetries, and organizing it in an expansion in derivatives and fields. Gapped excitations decay rapidly and do not affect long-distance behavior, whereas symmetry-protected modes such as hydrodynamic modes and Nambu-Goldstone (NG) modes dominate the infrared dynamics~\cite{ChaikinLubensky:1995}. A systematic method for constructing EFTs in phases with spontaneously broken symmetries is provided by the Callan–Coleman–Wess–Zumino coset construction~\cite{Coleman:1969sm,Callan:1969sn}.

In recent years, the notion of symmetry has been broadened beyond conventional global symmetries to include what are now collectively referred to as generalized symmetries~\cite{Gaiotto:2014kfa}
(see \cite{Bhardwaj:2023kri,Schafer-Nameki:2023jdn,Gomes:2023ahz,Brennan:2023mmt}
for recent reviews). In this framework, charged objects need not be pointlike, but may instead be
extended objects such as lines or surfaces. Many of the standard notions associated with ordinary symmetries, including Ward-Takahashi identities, coupling to background fields, gauging,
spontaneous symmetry breaking, and 't Hooft anomalies, admit natural generalizations.

Higher-form symmetries constitute a particularly important class of generalized symmetries. A notable example is that the photon can be reinterpreted as the Nambu-Goldstone mode associated with a spontaneously broken $\mathrm{U}(1)$ $1$-form symmetry~\cite{Gaiotto:2014kfa,Lake:2018dqm,Hofman:2018lfz}. A generalized coset construction that incorporates both ordinary (0-form) and higher-form symmetries was proposed in~\cite{Hidaka:2020ucc}, together with a unified counting rule for Nambu-Goldstone modes applicable to systems with both 0-form~\cite{Watanabe:2012hr,Hidaka:2012ym} and higher-form symmetries.

While most EFT studies focus on zero-temperature dynamics, thermal and nonequilibrium effects are crucial in many physical settings. At finite temperature, dissipation and noise become essential ingredients of the low-energy dynamics. Spontaneous symmetry breaking in such dissipative systems often gives rise to overdamped or diffusive collective modes rather than propagating NG modes~\cite{Minami:2015uzo,Hidaka:2019irz,Hongo:2019qhi}. 
A systematic framework for incorporating dissipation, noise, and symmetry constraints at finite temperature is provided by the Schwinger--Keldysh formalism \cite{Kamenev:2011,Crossley:2015evo,Glorioso:2017fpd,Glorioso:2016gsa,Haehl:2015pja,Haehl:2016pec,Haehl:2018lcu,Jensen:2018hse,Harder:2015nxa,glorioso2018lecturesnonequilibriumeffectivefield}, which enables a real-time description of nonequilibrium dynamics. In this framework, equilibrium constraints such as the fluctuation--dissipation relation are encoded via a discrete symmetry known as the dynamical Kubo--Martin--Schwinger (KMS) symmetry~\cite{Glorioso:2016gsa}.

In this work, we develop a systematic finite-temperature effective field theory for systems with spontaneously broken higher-form symmetries. As a concrete physical application, we formulate an EFT describing photon
dynamics in insulating media. The effective theory is constructed by combining the Schwinger--Keldysh formalism with the generalized coset construction, allowing for a systematic treatment of both conservative and dissipative effects. Imposing the dynamical KMS symmetry ensures consistency with the second law of thermodynamics and leads to general results such as the fluctuation--dissipation theorem and Onsager's reciprocal relations.

From this perspective, strong-to-weak symmetry breaking,
which has been discussed in the context of open quantum systems,
provides a useful conceptual framework for understanding diffusive
collective modes in dissipative systems~\cite{Lessa:2024gcw,Gu:2024wgc,Huang:2024rml}.
Closely related symmetry structures have also appeared in the Schwinger--Keldysh coset construction, where the doubling of the symmetry group 
$\mathcal G$ to
$\mathcal G \times \mathcal G$ plays a central role in formulating effective field theories for systems with dissipation and noise~\cite{Crossley:2015evo,Akyuz:2023lsm}. In this work, we revisit these ideas in the context of higher-form symmetries within the Schwinger--Keldysh framework and clarify how a diffusive shift symmetry emerges in the description of the unbroken phase.

The rest of the paper is organized as follows.
In section~\ref{sec:sk-review}, we review the general framework of Schwinger--Keldysh effective field theory. We also summarize the coset construction on the Schwinger--Keldysh contour for spontaneously broken 0-form symmetries, which serves as a prototype for the higher-form generalization.
In section~\ref{sec:eft-photon}, we construct a Schwinger--Keldysh effective field theory for photons based on spontaneously broken ${\rm U}(1)$ $1$-form symmetry. After identifying the appropriate low-energy degrees of freedom, we derive the most general effective Lagrangian in $D=3+1$ dimensions consistent with symmetries and power counting.
Section~\ref{sec:physical-properties} is devoted to the physical consequences of the dissipative photon effective field theory. We present an equivalent Langevin description, analyze the dispersion relations of photon modes, derive the fluctuation–dissipation relations, and construct the entropy current. We also discuss electromagnetic duality within the Schwinger--Keldysh framework.
In section~\ref{sec:unbroken}, we consider the effective field theory for an unbroken ${\rm U}(1)$ $1$-form symmetry and clarify the origin of diffusive symmetry.
In section~\ref{sec:strongweaksym}, we discuss the relation between the Schwinger--Keldysh symmetry structure and the notions of strong and weak symmetries in open quantum systems.
Finally, section~\ref{sec:summary} summarizes our results and outlines possible future directions.

Several technical points that would interrupt the flow of the main discussion
are collected in the appendices.
Specifically, appendix~\ref{app:ChernSimonsForm} summarizes the classification of
higher-form symmetry–invariant terms up to total derivatives,
appendix~\ref{sec:CME} discusses the details of the term representing the
chiral magnetic effect in the present framework,
and appendix~\ref{app:dkms} provides supplementary details on the dynamical KMS symmetry.

\paragraph{Notation and conventions.}
Before concluding the introduction, we briefly summarize some notations and conventions used in this work.
We work in $D$-dimensional Minkowski spacetime with $D=d+1$, where $d$ denotes the number of spatial dimensions, and adopt the metric
\begin{equation}
    \eta_{\mu\nu}
    = \mathrm{diag}(-1, \overbrace{+1,+1,\ldots,+1}^{d}) .
\end{equation}
Boldface symbols denote spatial vectors, e.g.,
\begin{equation}
 \bm A \coloneqq (A_1,\ldots,A_{d}), \qquad
 \bm\nabla \coloneqq (\partial_1,\ldots,\partial_{d}) .
\end{equation}
The totally antisymmetric tensor is normalized as
\begin{equation}
    \varepsilon_{01\cdots d} = +1, \qquad
    \varepsilon^{01\cdots d} = -1 .
\end{equation}
When discussing electromagnetic fields in $D=3+1$ dimensions,
the electric and magnetic fields are defined in terms of the field strength tensor as
\begin{equation}
    \bm E \coloneqq (F^{0i})
    = - \dot{\bm A} - \bm\nabla \changed{A^0}, \qquad
    \bm B \coloneqq (\varepsilon_{0ijk} F^{jk})
    = \bm\nabla \times \bm A .
\end{equation}

\section{Schwinger--Keldysh effective field theory}\label{sec:sk-review}

In this section, we review the formalism of Schwinger--Keldysh effective field theories, which provide a systematic framework for describing dissipative dynamics in quantum systems. The structure of the theory is strongly constrained by the unitarity of the underlying microscopic theory and by the assumption of thermal equilibrium of the initial state (sections~\ref{sec:SK_DKMS} and~\ref{sec:SKEFT}).
We also illustrate how the effective field theory should be constructed when a continuous 0-form global symmetry is spontaneously broken, as a prototype for higher-form symmetry breaking (section \ref{sec:SKEFT_SSB}). 
For further details, we refer the reader to a lecture note~\cite{glorioso2018lecturesnonequilibriumeffectivefield} and the original papers \cite{Crossley:2015evo,Glorioso:2017fpd,Akyuz:2023lsm}.

\subsection{Dynamical Kubo--Martin--Schwinger symmetry of generating functional}\label{sec:SK_DKMS}
To compute real-time expectation values and correlation functions in
nonequilibrium quantum systems, it is convenient to introduce a generating
functional $W[\phi_+,\phi_-]$ defined on a closed-time path,
\begin{align}
    e^{W[\phi_+,\phi_-]}
    =\Tr\left[\rho_0
    {\rm \tilde T}e^{-\im\int \rd^D x \mathcal O_-(t,\bm x)\phi_-(t,\bm x)}
    {\rm T} e^{\im\int \rd^D x \mathcal O_+(t,\bm x)\phi_+(t,\bm x)}
    \right],
\end{align}
where $\rho_0$ is the initial density matrix and $\rm T$ ($\rm\tilde T$) denotes (anti-)time ordering of operators.
For simplicity, we assume that sources $\phi_{\pm}$ are real and coupled to hermitian operators $\mathcal O_{\pm}$.
Note that we work in the interaction picture, treating the
source terms as interactions. 
Since the time integration starts from $t_i$ to $t_f$ and then comes back to $t_i$, the formalism is called closed-time path formalism.
By taking functional derivatives, $W[\phi_+,\phi_-]$ enables one to calculate expectation values and path-ordered (connected) correlation functions of operators coupled to the source.

The generating functional has several important properties.
From the unitarity of the time evolution, we can show
(i) $W[\phi_+=\phi, \phi_-=\phi]=0$, 
(ii) $W[\phi_+,\phi_-]^*=W[\phi_-,\phi_+]$, and
(iii) ${\rm Re} \ W[\phi_+,\phi_-]\leq 0$.
These properties hold for general initial density matrix $\rho_0$.
If the initial state is thermal equilibrium $\rho_0=e^{-\beta H}/Z_0$ with $Z_0=\Tr e^{-\beta H}$, the generating functional $W$ satisfies a relation called Kubo--Martin--Schwinger (KMS) condition:
\begin{align}
    e^{W[\phi_1,\phi_2]}
    =\frac{1}{Z_0}\Tr\left[e^{-\beta H}
    {\rm T}e^{\im\int \rd^D x \mathcal O_+(t-\im\theta,\bm x)\phi_+(t,\bm x)}
    {\rm \tilde T} e^{-\im\int \rd^D x \mathcal O_-(t+\im(\beta-\theta),\bm x)\phi_-(t,\bm x)}
    \right], \label{eq:KMS}
\end{align}
where the path order is reversed and the operators are shifted in the complex time plane.

Let us consider the case where the Hamiltonian has discrete symmetry $\Theta$ including time reversal $\mathcal T$ such as $\Theta=\{\mathcal{T, CT, PT, CPT}\}$, where $\mathcal{C}$ and $\mathcal{P}$ are charge conjugation and parity.
Under the antilinear transformation $\Theta$, the operator $\mathcal O$ transforms as
\begin{align}
    \Theta\mathcal O(t,\bm x)\Theta^{-1} = \eta_{\mathcal O}\mathcal O(-t,\eta \bm x),
\end{align}
where $\eta_{\mathcal O}=\pm 1$ and $\eta=\pm1$.
Using the property of the antilinear transformation $\Tr X = (\Tr \Theta X\Theta^{-1})^*$ and $[H,\Theta]=0$, the generating functional is written as
\begin{align}
    e^{W[\phi_1,\phi_2]}
    =\frac{1}{Z}\Tr\left[e^{-\beta H}
    {\rm \tilde T}e^{-\im\int \rd^D x \eta_{\mathcal O}\mathcal O_-(t+\im(\beta-\theta),\bm x)\phi_-(-t,\eta \bm x)}
    {\rm T} e^{\im\int \rd^D x \eta_{\mathcal O}\mathcal O_+(t-\im\theta,\bm x)\phi_+(-t,\eta \bm x)}
    \right].
\end{align}
On the right hand side, the path order is the same as that of $W[\phi_+,\phi_-]$, so the relation is a symmetry of $W[\phi_+,\phi_-]$:
\begin{subequations}
\begin{align}
    & W[\phi_+,\phi_-] = W[\tilde\phi_+,\tilde\phi_-],\\
    & \tilde\phi_+(t,\bm x) = \eta_{\phi}\phi_+(-t+\im\theta,\eta \bm x),\quad
    \tilde\phi_-(t,\bm x) = \eta_{\phi}\phi_-(-t-\im(\beta-\theta),\eta \bm x),
    \label{DKMS_transf_fields}
\end{align}
\label{DKMS_transf}
\end{subequations}
with $\eta_{\phi}=\eta_{\mathcal O}$.
This discrete, antilinear $\mathbb{Z}_2$ invariance
of the generating functional
is known as the dynamical Kubo--Martin--Schwinger (KMS) symmetry.

Throughout the discussion above, we have implicitly assumed that the discrete
symmetry $\Theta$ is not spontaneously broken. If $\Theta$ is spontaneously broken, the transformation~\eqref{DKMS_transf} is modified accordingly. See eq.~\eqref{DKMS_transf_SSB}, with details provided in
appendix~\ref{app:DKMS_SSB}.

\subsection{Effective field theory on the Schwinger--Keldysh contour}\label{sec:SKEFT}
In the path-integral formulation, the generating functional $W[\phi_1,\phi_2]$ is defined by functional integration over all the microscopic degrees of freedom $\psi$:
\begin{align}
    e^{W[\phi_1,\phi_2]}
    =\int \Dcal\psi_+ \Dcal\psi_- \rho_0[\psi_+,\psi_-]e^{\im I_0[\psi_+,\phi_+]-\im I_0[\psi_-,\phi_-]}, \quad
    \psi_+(t_f,\bm x)=\psi_-(t_f,\bm x),
\end{align}
where $I_0[\psi,\phi]$ denotes the action of the microscopic theory in the presence of an external source $\phi$.
The boundary condition at $t=t_f$ arises from inserting a complete set of
states at the final time.

When one is interested only in low-energy phenomena, the relevant degrees of
freedom $\chi$ are those appearing in hydrodynamic descriptions, such as
conserved densities and Nambu--Goldstone modes associated with spontaneously
broken continuous symmetries.
The generating functional can then be rewritten as a path integral over these
slow modes,
\begin{align}
    e^{W[\phi_+,\phi_-]}
    =\int \Dcal\chi_+ \Dcal\chi_- 
    e^{\im I_{\rm eff}[\chi_+,\phi_+,\chi_-,\phi_-]}, \quad
    \chi_+(t_f,\bm x)=\chi_-(t_f,\bm x).
\end{align}
The Schwinger--Keldysh effective action $I_{\rm eff}$ can be viewed as the generating functional in which both $\phi_{\pm}$ and $\chi_{\pm}$ act as sources for microscopic degrees of freedom, which have already been integrated out.
Consequently, the unitarity and the dynamical KMS conditions hold as well for $\phi_{\pm}$ and $\chi_{\pm}$ by a straightforward extension of the argument in the previous section.

Here, instead of repeating the similar arguments, we introduce another basis ($r$-$a$ basis) for the fields and sources
\begin{align}
    \chi_r = \frac{\chi_+ + \chi_-}{2}, \quad 
    \phi_r = \frac{\phi_+ + \phi_-}{2}, \quad
    \chi_a = \chi_+ - \chi_-,\quad   
    \phi_a = \phi_+ - \phi_-.
\end{align}
In this basis, the unitarity conditions are written as
\begin{subequations}
     \begin{align}
        \text{(i)} \quad  &I_{\rm eff}[\chi_r,\phi_r,\chi_a=0, \phi_a=0] = 0,
        \label{Unitarity1}\\
        \text{(ii)} \quad  &I_{\rm eff}^*[\chi_r,\phi_r,\chi_a,\phi_a] = - I_{\rm eff}[\chi_r,\phi_r,-\chi_a,-\phi_a],
        \label{Unitarity2}\\
        \text{(iii)} \quad  &{\rm Im} \ I_{\rm eff}[\chi_r,\phi_r,\chi_a,\phi_a]\geq 0.
        \label{Unitarity3}
    \end{align}
\end{subequations}
By an appropriate choice of total-derivative terms, these conditions may be
imposed locally on the Lagrangian density
$\mathcal L_{\rm eff}$ defined through 
$I_{\rm eff} = \int \rd^D x\, \mathcal L_{\rm eff}$.

The dynamical KMS transformation becomes simplified in the classical limit.
To track the scaling behavior as $\hbar\to0$, let us temporarily restore $\hbar$,
\begin{align}
    \chi_r \to \chi_r, \quad
    \chi_a \to \hbar\chi_a, \quad
    \beta,\theta \to \hbar \beta,\hbar\theta, \quad
    I_{\rm eff}\to \frac{1}{\hbar}I_{\rm eff}.
\end{align}
In the limit $\hbar\to 0$, the dynamical KMS transformation \eqref{DKMS_transf_fields} reduces to 
\begin{subequations}\begin{align}
    \tilde \chi_r(t,\bm x) &= \eta_{\chi}\chi_r(-t,\eta \bm x),\quad
    \tilde \chi_a(t,\bm x) = \eta_{\chi}\chi_a(-t,\eta \bm x) + \im\beta\eta_{\chi}\dot\chi_r(-t,\eta \bm x), \\
    \tilde \phi_r(t,\bm x) &= \eta_{\phi}\phi_r(-t,\eta \bm x),\quad
    \tilde \phi_a(t,\bm x) = \eta_{\phi}\phi_a(-t,\eta \bm x) + \im\beta\eta_{\phi}\dot\phi_r(-t,\eta \bm x),
\end{align}\label{DKMStrsf0form}\end{subequations}
where $\eta_{\chi,\phi}=\pm 1$ are $\mathbb{Z}_2$ charges of $\chi,\phi$ under $\Theta$, 
and and $\dot f \equiv \partial_t f$.
This transformation leaves the effective action invariant in classical limit. Let us introduce the operator to substitute dynamical-KMS-transformed fields,
\begin{equation}
    \mathrm{KMS}_{\Theta}\{\Ocal[\chi_r,\phi_r,\chi_a, \phi_a]\}
    \coloneqq
    \Ocal [\tilde\chi_r,\tilde\phi_r,\tilde\chi_a,\tilde\phi_a].
\end{equation}
Then the invariance of $I_{\rm eff}$ in terms of the Lagrangian density $\Lcal_{\rm eff}$ is written as
\begin{equation}
    \int\rd^D x\
        \mathrm{KMS}_{\Theta}
        \{\Lcal_{\rm eff}[\chi_r,\phi_r,\chi_a, \phi_a]\}
    =  \int\rd^D x\
        \Lcal_{\rm eff}[\chi_r,\phi_r,\chi_a, \phi_a].
\end{equation}
Since the transformation \eqref{DKMStrsf0form} involves a coordinate transformation, to obtain a local relation between the Lagrangian densities, one has to perform a change of integration variables $(-t,\eta \bm x)\to (t,\bm x)$.
We denote this operation by $\mathrm{T}_\Theta$, which acts on fields and
derivatives as
\begin{equation}\begin{aligned}
    &\mathrm{T}_\Theta[\chi_{a,r}(t,\bm x)]
    = \chi_{a,r}(-t,\eta \bm x),
    &&\mathrm{T}_\Theta[\phi_{a,r}(t,\bm x)]
    = \phi_{a,r}(-t,\eta \bm x), \\
    &\mathrm{T}_\Theta[\p_t]
    = -\p_t,
    &&\mathrm{T}_\Theta[\bm\nabla_x]
    = \eta\bm\nabla_x.
\end{aligned}\end{equation}
Combining these relations, we arrive at the dynamical KMS condition for Lagrangian densities,\footnote{
If $\Theta$ is spontaneously broken, the condition \eqref{DKMS_Leff} is modified
accordingly (see eq.~\eqref{DKMS_transf_SSB}).
}
\begin{equation}
    \mathrm{T}_\Theta\cdot\mathrm{KMS}_{\Theta}
        \{\Lcal_{\rm eff}[\chi_r,\phi_r,\chi_a, \phi_a]\}
    = \Lcal_{\rm eff}[\chi_r,\phi_r,\chi_a, \phi_a]
        + \p_\mu V^\mu [\chi_r,\phi_r,\chi_a, \phi_a].
    \label{DKMS_Leff}
\end{equation}

Finally, we note that the dynamical KMS transformation mixes different orders
in the derivative expansion, as is evident from \eqref{DKMStrsf0form}.
This requires that, in the power-counting scheme, the number of $a$-type fields and the number of time derivatives should be counted equally.

\subsection{Schwinger--Keldysh effective field theory with broken global symmetry}\label{sec:SKEFT_SSB}

In the previous sections, we discussed general constraints on the generating
functional $W[\phi_+,\phi_-]$ and on the effective action $I_{\rm eff}[\chi_r,\phi_r,\chi_a,\phi_a]$ arising from unitarity and the dynamical KMS symmetry. 
When the microscopic theory has a continuous global symmetry $\mathcal G$, and in particular the symmetry is spontaneous broken to $\mathcal H$ in the ground state, 
the effective field theory is further constrained by both the symmetry and its
pattern of breaking.

In this section, we briefly review the symmetry in the Schwinger--Keldysh path integral and the coset construction when the symmetry is spontaneously broken, following \cite{Akyuz:2023lsm}. As a concrete illustration, we explain how to construct a Schwinger--Keldysh effective field theory for a system with a spontaneously broken global $\rm U(1)$ 0-form symmetry.

\subsubsection{Schwinger--Keldysh coset construction}\label{sec:SKcoset}

In the Schwinger--Keldysh path-integral formulation,
a global symmetry $\mathcal G$ is naturally doubled to $\mathcal G_+\times\mathcal G_-$,
where $\mathcal G_+$ and $\mathcal G_-$ act independently on fields defined
on the forward and backward time contours, respectively.
This doubling can be understood by recalling that the Schwinger--Keldysh
effective action governs the real-time evolution of the density matrix.

At the microscopic level, the density matrix evolves according to 
$\dot\rho=\mathcal L[\rho]=-\im[H,\rho]$, where $\mathcal L$ is 
a Liouvillian operator.
The global symmetry implies that the Hamiltonian is invariant $U^{\dagger}HU=H$ with $U(g)$ being a representation of $g\in \mathcal G$ in the Hilbert space.
Then, the Liouvillian has a doubled symmetry $\mathcal G_+\times \mathcal G_-$, i.e., $\mathcal L[\rho] = U_+^{\dagger}\mathcal L\left[U_+ \rho U_-^{\dagger}\right]U_-$ for an arbitrary $\rho$.
For a finite time interval, we have
\begin{align}
    \rho(t_2) = e^{(t_2-t_1)\mathcal L}[\rho(t_1)]
    = U_+^{\dagger}\left(e^{(t_2-t_1)\mathcal L}[U_+\rho(t_1)U_-^{\dagger}]\right)U_-.
\end{align}
We require that the time evolution generated by the Schwinger--Keldysh effective
action, represented as a linear map $\mathcal V_{\rm eff}(t)[\cdot]$, respect the same symmetry,
\begin{align}
    \rho_R(t_2) = \mathcal V_{\rm eff}(t_2-t_1)[\rho_R(t_1)]
    = U_+^{\dagger}\left(\mathcal V_{\rm eff}(t_2-t_1)[U_+\rho_R(t_1)U_-^{\dagger}]\right)U_-,
\end{align}
where $\rho_R$ denotes the reduced density matrix describing the low-energy
degrees of freedom.
In the path-integral language, this implies that the fields on the forward and
backward contours transform independently under $\mathcal G_+$ and
$\mathcal G_-$.

In the context of open quantum systems, such a doubled symmetry 
is referred to as a {\it strong symmetry}, while the diagonal subgroup, i.e., $U_+ = U_-$, corresponds to a {\it weak symmetry}.
We will discuss the parallel between strong and weak symmetry structures in the Schwinger–Keldysh EFT and in open quantum systems in section~\ref{sec:strongweaksym}.

The initial density matrix $\rho(t_i)$ and the continuity condition for fields at $t=t_f$ are boundary conditions that may break this symmetry.
The latter does not affect how we construct an effective action as we impose the boundary condition at $t=t_f$ after writing down an effective Lagrangian.\footnote{
The integration over microscopic gapped modes with continuity condition at $t=t_f$ generally implies that only
\changed{%
the diagonal subgroup $\mathcal G_r$
}%
remains as a symmetry of the effective action, rather than the full doubled symmetry.
For example, discrete symmetries such as $\mathcal{C, P, CP}$ are not doubled.
However, we expect our effective field theory to retain doubled symmetries for continuous global ones because the information about the conserved charge density is still encoded in the effective action.
}

At first sight, the initial equilibrium density matrix $\rho\propto e^{-\beta H}$ 
appears to be always invariant under the symmetry transformation $U\rho U^{\dagger}=\rho$, i.e.\ %
\changed{%
diagonal subgroup $\mathcal G_{\it r}\subset\mathcal G_+\times \mathcal G_-$.
}%
However, when the symmetry is broken in the vacuum, we need to introduce an infinitesimal breaking term to orient the vacuum configuration to a particular direction.
Because of this infinitesimal breaking term, the density matrix is invariant only under
\changed{%
the diagonal subgroup $\changed{\mathcal H_{\it r}}\subset \mathcal H_+\times \mathcal H_-$.
}%

Given
\changed{%
the symmetry breaking pattern $\mathcal G_+\times \mathcal G_-\to \mathcal H_{\it r}$,
}%
the coset manifold is parametrized by variables associated with the broken generators, from which the Maurer--Cartan $1$-form is constructed.
\changed{%
Let us denote the broken generators for $\mathcal G\to\mathcal H$ as $X^A$
and those for unbroken ones as $T^A$,
where $A$ is a generator index.
Upon doubling the symmetry to 
$\mathcal G_+ \times \mathcal G_-$, 
we introduce the generators 
$X_\pm^A$ and $T_\pm^A$ acting on the forward and backward branches of the CTP contour.
Then the coset variables are $\pi_r^A$ for $\ X_r^A=X_+^A + X_-^A$, $\ \pi_a^A$ for $X_a^A=X_+^A-X_-^A\ $ and $\varphi_a^A$ for $\ T_a^A=T_+^A-T_-^A\ $.
The coset manifold is parametrized as 
\begin{equation}
\Omega=e^{\im\pi_r^A X_r^A}e^{\im\pi_a^A X_a^A}e^{\im\varphi_a^A T_a^A}.
\end{equation}
The diagonal generators
$\ T_r^A = T_+^A + T_-^A\ $ 
remain unbroken and are therefore not included in the coset parametrization.
Under a global transformation 
$g_+ g_-\in \mathcal G_+\times\mathcal G_-$,
the coset element transforms as
\begin{equation}
g_+ g_-\Omega(\pi,\varphi)
=\Omega(\pi',\varphi')\,h_r ,
\end{equation}
where $h_r\in\mathcal H_r$.
This defines the nonlinear transformation of the fields
$\pi_{a,r}^A$ and $\varphi_a^A$.
}%

At the level of the Schwinger--Keldysh effective field theory, however, it is convenient to introduce a field variable associated with
\changed{%
the diagonal generator $T_r^A$,
even though $\mathcal H_r$ is not spontaneously broken.
}%
We therefore enlarge the coset parametrization by introducing
\changed{%
a field $\varphi_r^A$ through
\begin{equation}
    \tilde\Omega = \Omega\, e^{\im\varphi_r^A T_r^A},
\end{equation}
}%
and subsequently impose a gauge symmetry to eliminate unphysical degrees of freedom.
\changed{%
When $\mathcal G$ is Abelian,
the fields transform by constant shifts:
\begin{equation}
    \pi_{a,r}^A
    \longmapsto
    {\pi'}_{a,r}^A
    = \pi_{a,r}^A + c_{a,r}^A,
    \qquad
    \varphi_{a,r}^A
    \longmapsto
    {\varphi'}_{a,r}^A
    = \varphi_{a,r}^A + b_{a,r}^A.
\end{equation}
}

\changed{%
The Maurer--Cartan $1$-form is then defined as
\begin{align}
    \omega \coloneqq -\im\tilde\Omega^{-1}\rd\tilde\Omega
    = (D\pi_r^A)X_r^A + (D\pi_a^A)X_a^A + (D\varphi_a^A)T_a^A + (D\varphi_r^A)T_r^A.
\end{align}
}%
which furnishes the building blocks of the effective action.
\changed{%
In the Abelian case,
the covariant derivative $D$ reduces to the ordinary exterior derivative,\footnote{
\changed{%
For non-Abelian symmetries, the expression of covariant derivatives can be computed using the Lie algebra (see, e.g., eq.~(69) in \cite{PhysRevX.4.031057}).
}%
}
\begin{equation}
D\pi_{a,r}^A = \rd \pi_{a,r}^A,
\quad
D\varphi_{a,r}^A = \rd\varphi_{a,r}^A,
\end{equation}
Since $\mathcal H_r$ remains unbroken,
}%
we require the invariance under local transformations, referred to as the {\it diffusive symmetry}~\cite{glorioso2018lecturesnonequilibriumeffectivefield},
\changed{
\begin{align}
    \tilde\Omega\to \tilde\Omega e^{\im \lambda^A(\bm x)T_r^A},
\end{align}
where $\lambda^A (\bm x)$ is
}%
an arbitrary function of spatial coordinates. 
This invariance implies that
\changed{%
$\varphi_r^A$ can enter the effective action only through the combination
$\rho_r^A \coloneqq D_t \varphi_r^A$,
}%
which is naturally interpreted as the conserved density associated
\changed{%
with $\mathcal H_r$.
}%
We give a more detailed discussion on the diffusive symmetry and its relation to strong/weak symmetries in section~\ref{sec:unbroken}.

\subsubsection{Example: Schwinger--Keldysh effective action for \texorpdfstring{${\rm U}(1)_+\times {\rm U}(1)_-\to 1$}{U(1)\_+ × U(1)\_- to 1}}
As a simple illustrative example relevant to our study, we show how to construct the Schwinger--Keldysh effective action for the symmetry breaking ${\rm U}(1)_+\times {\rm U}(1)_-\to 1$.
In this case, the coset manifold $\Omega$ is parametrized by the phonon fields $\pi_{a,r}$ as
\changed{%
\begin{align}
    \Omega =
    e^{\im(\pi_r + \pi_a)Q_+}\,
    e^{\im(\pi_r-\pi_a)Q_-},
\end{align}
where $Q_\pm$ are the total charges 
associated with ${\rm U}(1)_\pm$.
Indices for generators are omitted because ${\rm U}(1)_\pm$ has only one generator, respectively.
}%
Under $e^{\im (\lambda_r+\lambda_a)Q}\times e^{\im (\lambda_r-\lambda_a)Q}\in {\rm U}(1)_+\times {\rm U}(1)_-$, phonon fields are shifted by constants $\pi_{a,r}\mapsto\pi_{a,r}+\lambda_{a,r}$.
From the coset parametrization $\Omega$, the Maurer--Cartan $1$-form,
\changed{%
$\omega \coloneqq -\im\Omega^{\dagger}\rd\Omega$, 
is given by
\begin{align}
\omega
    = \rd\pi_r (Q_+ + Q_-) + \rd\pi_a (Q_+ - Q_-).
\end{align}
}%
The symmetry is fully broken, so the effective action is constructed
from $\partial_{\mu}\pi_r$ and $\partial_{\mu}\pi_a$ without imposing the diffusive symmetry.\footnote{
\changed{%
We discuss the origin of the diffusive symmetry
in section~\ref{sec:origin-diffusive-symmetry}.
}}

Let us construct the effective action respecting the unitarity and rotational symmetry.
We adopt a power counting scheme $N=m+n$, where $m$ counts the order of $\pi_a$ field and $n$ counts the derivatives in time and space.
We list all the terms $\mathcal O_{(m,n)}[\pi_a,\pi_r]$ with $m+n\leq 4$;
\begin{subequations}\begin{align}
    \mathcal O_{(1,1)}[\pi_a,\pi_r] &: 
    \pi_a \p_t\pi_r, \\
    \mathcal O_{(1,2)}[\pi_a,\pi_r] &: 
    \p_t\pi_a \p_t\pi_r, \
    \bm\nabla\pi_a\cdot\bm\nabla\pi_r, \\
    \mathcal O_{(1,3)}[\pi_a,\pi_r] &: 
    \p_t\pi_a \partial_t^2\pi_r, \
    \p_t\pi_a \bm\nabla^2\pi_r, \
    \p_t\pi_a (\p_t\pi_r)^2, \
    \p_t\pi_a (\bm\nabla\pi_r)^2, \
    \p_t\pi_r\bm\nabla\pi_a\cdot \bm\nabla\pi_r,  \\
    \mathcal O_{(2,2)}[\pi_a,\pi_r] &: 
    \p_t\pi_a\partial_t\pi_a, \
    \bm\nabla\pi_a\cdot\bm\nabla\pi_a,
\end{align}\end{subequations}
where $m=0$ is not allowed because of the unitarity condition \eqref{Unitarity1}.
The term in $\Ocal_{(1,1)}$ is special in that it cannot be written solely in terms of $\partial_\mu \pi_{a,r}$.
Instead, the symmetry is realized only up to a total derivative. 
Such terms are of Wess--Zumino--Witten type and reflect a non-gaugeable realization of the symmetry.
Imposing the dynamical KMS symmetry $I_{\rm eff}[\pi_r, \pi_a] = I_{\rm eff}[\tilde\pi_r, \tilde\pi_a]$ with
\begin{align}
    \tilde\pi_r(t,\bm x) = \eta_{\pi}\pi_r(-t,\eta \bm x),\quad
    \tilde\pi_a(t,\bm x) = \eta_{\pi}\pi_a(-t,\eta \bm x)+\im\beta\eta_{\pi}\p_{-t}\pi_r(-t,\eta \bm x),
\end{align}
the building blocks of $\mathcal L_{\rm eff}$ for $\eta_{\pi}=1$ are
\begin{align}
    \p_t\pi_a \p_t\pi_r, \
    \bm\nabla\pi_a\cdot\bm\nabla\pi_r,\
    \p_t\pi_a \p_t(\pi_a + \im\beta\p_t\pi_r),\
    \bm\nabla\pi_a \cdot \bm\nabla(\pi_a + \im\beta\p_t\pi_r)
\end{align}
and those for $\eta_{\pi}=-1$ are
\begin{equation}\begin{split}
    &\p_t\pi_a \p_t\pi_r, \
    \bm\nabla\pi_a\cdot\bm\nabla\pi_r,\
    \p_t\pi_a \p_t(\pi_a + \im\beta\p_t\pi_r),\
    \bm\nabla\pi_a \cdot \bm\nabla(\pi_a + \im\beta\p_t\pi_r),\\
   &\p_t\pi_a (\p_t\pi_r)^2,\
    (\p_t\pi_a\bm\nabla\pi_r + 2\p_t\pi_r\bm\nabla\pi_a)\cdot\bm\nabla\pi_r.
\end{split}\end{equation}
The effective Lagrangian\footnote{
    We absorbed the coefficient of the $\dot\pi_a\dot\pi_r$ term by time-rescaling. This coefficient is always positive in order for the ground state to be stable, i.e., $\Im\omega(k)<0$.
}
is their linear combination. In the case of $\eta_\pi=+1$,
\begin{align}
    \label{effLag_0form_1}
    \Lcal_{\rm eff}
    =   \dot\pi_a\dot\pi_r
        - c_s^2\bm\nabla\pi_a\cdot\bm\nabla\pi_r
        - \tau\dot\pi_a\ddot\pi_r
        - \sigma\bm\nabla\pi_a\cdot\bm\nabla\dot\pi_r
        + \frac{\im}{\beta}\tau\dot\pi_a^2
        + \frac{\im}{\beta} \sigma(\bm\nabla\pi_a)^2,
\end{align}
and for $\eta_\pi=-1$,
\begin{equation}\begin{split}
    \label{effLag_0form_2}
    \Lcal_{\rm eff}
    =   \dot\pi_a\dot\pi_r
        - c_s^2\bm\nabla\pi_a\cdot\bm\nabla\pi_r
        - \tau\dot\pi_a\ddot\pi_r
        - \sigma\bm\nabla\pi_a\cdot\bm\nabla\dot\pi_r
        + \frac{\im}{\beta}\tau\dot\pi_a^2
        + \frac{\im}{\beta} \sigma(\bm\nabla\pi_a)^2\\
        + \frac{1}{2}g_1\dot\pi_a\dot\pi_r^2
        + \frac{1}{2}g_2(\dot\pi_a\bm\nabla\pi_r+2\dot\pi_r\bm\nabla\pi_a)\cdot\bm\nabla\pi_r.
\end{split}\end{equation}
From the unitarity condition \eqref{Unitarity2}, imaginary units are introduced so that the odd-$m$ terms are real while the even-$m$ terms are purely imaginary. The unitarity condition \eqref{Unitarity3} restricts some of the coupling constants as,
\begin{equation}
    \tau \ge0,\
    \sigma \ge 0.
\end{equation}
The Lagrangians \eqref{effLag_0form_1} and \eqref{effLag_0form_2} are general up to $m+n\le4$ respecting the rotational, internal $\rm U(1)$, and dynamical KMS symmetries and all of the three unitarity conditions.

As a final remark, we consider the dispersion relation of the phonon modes.
The dispersion relation is determined by the quadratic (free) part of the
effective Lagrangian $\mathcal L_{\rm eff}^{(0)}$, which is the same for
$\eta_{\pi}=\pm1$:
\begin{align}
        \Lcal_{\rm eff}^{(0)}
    &=   \pi_a\left[-\partial_t^2
        + c_s^2\bm\nabla^2
        + \tau\partial_t^3
        + \sigma\bm\nabla^2\partial_t\right]\pi_r
        - \frac{\im}{\beta}\pi_a\left[\tau\partial_t^2 + \sigma\bm\nabla^2\right]\pi_a,
\end{align}
where we have performed integration by parts.
Using the invariance of $\langle\pi_r(t,\bm x)\rangle_0$ with respect to the change of path integral variable $\pi_r(t,\bm x)\to \pi_r(t,\bm x)+\delta\pi_r(t,\bm x)$, one finds that the retarded phonon propagator satisfies
\begin{align}
    \left[-\partial_t^2
        + c_s^2\bm\nabla^2
        + \tau\partial_t^3
        + \sigma\bm\nabla^2\partial_t\right]
        \langle\pi_r(t,\bm x)\pi_a(t',{\bm x}')\rangle_0 = \im\delta(t-t')\delta^{d-1}(\bm x- {\bm x}').
\end{align}
Here $\langle\cdots\rangle_0$ denotes the expectation value in the free theory $\Lcal_{\rm eff}^{(0)}$.

Applying Fourier transformation, adopting the convention
$f(t,\bm x)=\int_{\omega,\bm k} e^{\im \bm k\cdot\bm x-\im\omega t} f(\omega,\bm k)$,
the dispersion relation is determined by the pole of the propagator,
\begin{equation}
    0=
    \omega^2
    - c_s^2 k^2
    + \im\tau\omega^3
    + \im \sigma \omega k^2.
\end{equation}
In the long-wavelength limit $k\to 0$, this yields a $k^2$ correction to the linear dispersion relation:
\begin{equation}
    \omega(k)
    =
    \pm c_s k
    - \frac{\im}{2}\qty(\tau c_s^2 + \sigma)k^2
    + \Ocal(k^3).
\end{equation}
This result shows that the Nambu--Goldstone mode exhibits a linear dispersion
relation and acquires a finite lifetime at nonzero temperature due to
dissipative interactions with the thermal bath.
The dispersion relation also admits another solution,
$\omega(k=0)=\im/\tau$, which is gapped and corresponds to an unstable mode.
This mode is an artifact of truncating the derivative expansion at finite
order and should therefore be discarded within the regime of validity of the
effective theory.

\section{Effective field theory for photons from higher-form symmetries}\label{sec:eft-photon}

In this section, we discuss a Schwinger--Keldysh effective field theory for photons based on higher-form symmetries. After identifying the relevant low-energy degrees of freedom associated with a spontaneously broken $\mathrm{U}(1)$ $1$-form symmetry, we construct the corresponding Schwinger--Keldysh effective action.

\changed{%
We discuss a $1$-form symmetry associated with an on-shell conservation law rather than an off-shell identity.
In electromagnetism,
the $\mathrm U(1)$ electric $1$-form symmetry is associated with the Maxwell equation of motion in the absence of charged matter, as an on-shell conservation law,
while the $\mathrm U(1)$ magnetic $1$-form symmetry is associated with the Bianchi identity as an off-shell conservation law.
Therefore,
in the case of electromagnetism,
the $\mathrm{U}(1)$ $1$-form symmetry in our EFT
corresponds to the electric $1$-form symmetry.
Since the $\mathrm U(1)$ electric $1$-form symmetry will be explicitly broken by the existence of charged matter,
the situation of our EFT can be realized
in electromagnetism in `insulating' media,
namely media without electrically charged excitations.
}

Focusing on the case of $D=3+1$ dimensions, we derive the most general effective Lagrangian consistent with symmetry, power counting, unitarity, and the dynamical KMS condition. We also discuss the appearance of parity-violating effects such as the chiral magnetic effect.

\changed{%
In the IR,
hydrodynamic modes arising from energy--momentum conservation may also appear.
To highlight the effects of higher-form SSB on the IR dynamics
and for simplicity,
we consider a regime
where the electromagnetic and fluid sectors effectively decouple.
In this work,
the fluid sector is therefore not included in our EFT.%
}%
\footnote{\changed{%
Although this assumption does not hold in all systems,
it can be reasonable in many realistic situations.
For example,
many paramagnetic media exhibit weak magnetic responses,
and electric dipole and higher multipole excitations in insulators are efficiently screened.
Therefore,
the hydrodynamic modes typically appear at even longer time and length scales
than the electromagnetic sector,
so that their influence on the electromagnetic dynamics is suppressed.
}}%

\subsection{Dynamical degrees of freedom}

To construct an EFT for photons based on symmetry, the first step is to identify the appropriate dynamical degrees of freedom.  
To this end, we employ an extension of the coset construction,
originally developed for spontaneously broken 0-form symmetries
\cite{Coleman:1969sm,Callan:1969sn},
to the case of higher-form symmetries~\cite{Hidaka:2020ucc}.

We begin by recalling that, for an ordinary 0-form symmetry that is spontaneously broken, the coset variable serves as the Nambu--Goldstone mode parameterizing the broken symmetry directions. In this case, the Maurer--Cartan form constructed from the coset variable plays a central role 
in building the low-energy effective theory, as reviewed in the previous section.

A natural generalization of the MC form exists for higher-form symmetries as well. 
Suppose a $\rm U(1)$ $p$-form symmetry is spontaneously broken, ${\rm U}(1)^{[p]} \to 1$. 
The corresponding coset variable can be written as 
\begin{equation}
  W (C_p) = e^{\im \int_{C_p} A^{(p)}} ,
\end{equation}
where $C_p$ is a closed $p$-dimensional subspace. The $p$-form field $A^{(p)}$ appearing here is not unique: it enjoys a gauge redundancy,
\begin{equation}
  A^{(p)} \ \longmapsto\   A^{(p)} + \rd \alpha^{(p-1)} ,
  \label{GaugeTRSF}
\end{equation}
with $\alpha^{(p-1)}$ a $(p-1)$-form gauge parameter.  
This redundancy reflects the fact that $A^{(p)}$ only appears integrated over the closed subspace $C_p$.

We define the generalized Maurer--Cartan form $F^{(p+1)}$, a $(p+1)$-form, through\footnote{
For $p=0$, the left-hand side of eq.~\eqref{eq:g-mc} is understood to be path-ordered.
}
\begin{equation}
e^{\im \int_{X_{p+1}} F^{(p+1)}}
\coloneqq W^\dag (C'_p)\, W (C_p),
\label{eq:g-mc}
\end{equation}
where $X_{p+1}$ is a manifold with boundary $\p X_{p+1} = C_p \cup (-C'_p)$. 
For the 0-form case, taking the exterior derivative of both sides of eq.~\eqref{eq:g-mc} 
recovers the usual MC form:
\begin{equation}
  F^{(1)} = - \im W^\dagger (x)\, \rd W(x),
\end{equation}
with $W(x) = e^{\im A^{(0)}_A X^A}$. 
By taking the exterior derivative of both sides again, the Maurer--Cartan equation is obtained:
\begin{equation}
  \rd F^{(1)} + \im F^{(1)} \wedge F^{(1)} = 0,
\end{equation}
which ensures that the parallel transport is independent of the interpolating manifold $X_1$.
For $p \ge 1$, the symmetry is abelian and the generalized MC form reduces to
\begin{equation}
  F^{(p+1)} = \rd A^{(p)}.
\end{equation}
Since the right-hand side does not depend on the choice of $X_{p+1}$, 
$F^{(p+1)}$ must satisfy
\begin{equation}
  \rd F^{(p+1)} = 0.
  \label{Bianchi}
\end{equation}
Thus $F^{(p+1)}$ is a closed $(p+1)$-form and serves as the conserved current 
of the dual $(D-p-2)$-form symmetry. 
This dual symmetry is emergent in the broken phase.

\subsection{Schwinger--Keldysh effective action for \texorpdfstring{$\rm{U}(1)_+^{[\it p]}\times\rm{U}(1)_-^{[\it p]}\to1$}{U(1)\^[p]\_- × U(1)\^[p]\_+}}

In the previous section, we saw that the gauge field $A^{(p)}$ parametrizes the coset associated with the spontaneous breaking of a ${\rm U}(1)$ $p$-form symmetry, and thus serves as the relevant low-energy degree of freedom at zero temperature. In what follows, we discuss how to construct the corresponding Schwinger--Keldysh effective field theory.

The group structure of higher-form symmetry which produces gapless modes are tightly restricted. For $p\ge 1$, the higher-form symmetries are always abelian. Moreover, the gapless Nambu--Goldstone modes emerge only from continuous symmetry. It implies we can neglect discrete symmetries on determining the gapless degrees of freedom. If restricted to compact symmetries, the possible group structures are products of ${\rm U}(1)$ symmetries.
For this reason, here we discuss  spontaneous breaking of a ${\rm U}(1)$ $p$-form symmetry, ${\rm U}(1)^{[p]}$.

As discussed in section~\ref{sec:SKcoset}, the global symmetry \( \mathcal{G} \) is doubled to \( \mathcal{G}_+ \times \mathcal{G}_- \) on the closed-time path. In the Schwinger--Keldysh effective theory for 
\(
\mathrm{U}(1)_+^{[p]} \times \mathrm{U}(1)_-^{[p]} \to 1,
\)
the coset variable is also doubled as
\begin{equation}
    W_+(C_p)=e^{\im\int_{C_p}A_+^{(p)}},
    \qquad
    W_-(C_p)=e^{\im\int_{C_p}A_-^{(p)}},
\end{equation}
where \( A_\pm^{(p)}(t, x) \) parametrize the broken symmetry directions for \(\mathrm{U}(1)_+^{[p]}\) and
\(\mathrm{U}(1)_-^{[p]}\), respectively,
i.e., they serve as Nambu--Goldstone fields.
Hereafter, the superscript \( (p) \) will sometimes be omitted for brevity. 
The transformation rules of \( A_\pm \) are given by
\begin{equation}
    \rm{U}(1)^{[\it p]}_+: \left\{\begin{aligned}
        A_+   &\longmapsto A_+ + \Lambda_+\\
        A_-   &\longmapsto A_-
    \end{aligned}\right.
    \quad
    \rm{U}(1)^{[\it p]}_-: \left\{\begin{aligned}
        A_+   &\longmapsto A_+\\
        A_-   &\longmapsto A_- + \Lambda_-\\
    \end{aligned}\right.
    \qquad
    \text{where}\;\; \rd \Lambda_\pm^{(\it p)}=0.
\end{equation}
Now we move to $r$-$a$ basis:
\begin{equation}
    A \coloneqq \frac{A_++A_-}{2},
    \qquad
    a \coloneqq A_+ - A_-.
\end{equation}
Due to the Abelian nature of $\rm U(1)$,
\begin{equation}
    {\rm U}(1)_+^{[p]}\times{\rm U(1)}_-^{[p]}
    \cong
    {\rm U}(1)_{r}^{[p]}\times{\rm U}(1)_{a}^{[p]},
\end{equation}
where the coset variables of $\rm U(1)_{\it r}^{[\it p]}$ and $\rm U(1)_{\it a}^{[\it p]}$ are, respectively,
\begin{equation}
    W(C_p)=e^{\im\int_{C_p}A^{(p)}},
    \qquad
    w(C_p)=e^{\im\int_{C_p}a^{(p)}}.
    \label{eq:WilsonLoops}
\end{equation}
Since the Nambu--Goldstone fields $(A,a)$ are re-parametrization of $(A_+,A_-)$, their transformation rules are given by
\begin{equation}
    \rm{U}(1)^{[\it p]}_{\it r}: \left\{\begin{aligned}
        A   &\longmapsto A + \Lambda\\
        a   &\longmapsto a
    \end{aligned}\right.
    \quad
    \rm{U}(1)^{[\it p]}_{\it a}: \left\{\begin{aligned}
        A   &\longmapsto A\\
        a   &\longmapsto a + \lambda
    \end{aligned}\right.
    \quad
    \text{where}\;\;
    \left\{\begin{aligned}
        \rd \Lambda &= 0\\
        \rd \lambda &= 0
    \end{aligned}\right. .
\end{equation}
These $p$-form fields also possess gauge redundancies, as discussed in eq.~\eqref{GaugeTRSF}. 
These gauge fields provide a complete description of the directions associated with the spontaneous symmetry breaking
\begin{equation}
    \mathrm{U}(1)_{r}^{[p]} \times \mathrm{U}(1)_{a}^{[p]} \longrightarrow 1,
\end{equation}
and serve as the effective low-energy degrees of freedom of the EFT.

The Ward--Takahashi identity plays a fundamental role in the higher-form symmetries. In order to confirm the $\rm U(1)_{\it r}^{[\it p]}\times\rm U(1)_{\it a}^{[\it p]}$ symmetry is realized in the present theory, we derive the Ward--Takahashi identity from the transformation rule introduced in an ad hoc manner. 
Let us consider a variation of the action under 
a generic shift $A \longmapsto A + \Lambda$,
\begin{equation}
\delta I_{\rm eff} = \int \Lambda \wedge \frac{\delta I_{\rm eff}}{\delta A}.
\label{eq:variation}
\end{equation}
When $\Lambda$ is a closed form, the variation should be a total derivative, because we assume that this theory respects ${\rm U}(1)_r^{[p]}$ symmetry.
This requires that the variation should also be written in the form
\begin{equation}
\delta I_{\rm eff} = \int \Lambda \wedge \rd \star j.
\label{eq:variation_symmetry}
\end{equation}
Since $\Lambda$ is an arbitrary $p$-form,
eqs.~\eqref{eq:variation} and \eqref{eq:variation_symmetry} implies that the equation of motion is written as a conservation law
\begin{equation}
    (0 =)\
    \fdv{I_{\rm eff}}{A}
    = \rd\star j[a,A]
    \qquad
    \text{(on-shell).}
\end{equation}
Exactly the same logic holds for the ${\rm U}(1)_{a}^{[p]}$ symmetry to produce another conservation law,
\begin{equation}
    (0 =)\
    \fdv{I_{\rm eff}}{a}
    =: \rd\star J[a,A]
    \qquad
    \text{(on-shell).}
    \label{ConservedCurr}
\end{equation}
Now we can derive the Ward--Takahashi identities:
\begin{subequations}
\begin{equation}\begin{split}
    \ev{W[C_p]\exp(-\im\theta\int_\Mcal\star j)}
    &= \int\Dcal A\, \Dcal a\;W[C_p]\exp(-\im\theta\int_\mathcal{B}\fdv{I_{\rm eff}}{A})e^{\im I_{\rm eff}}\\
    &= \int\Dcal A\, \Dcal a\;e^{\im I_{\rm eff}}\exp(\theta\int_\mathcal{B}\fdv{A})W[C_p]\\
    &= \exp(\im\theta\;\mathrm{Link}[\Mcal,C_p])\ev{W[C_p]},
\end{split}\end{equation}\label{eq:wt-r}
\begin{equation}
    \ev{w[C_p]\exp(-\im\theta\int_\Mcal\star J)}
    = \exp(\im\theta\;\mathrm{Link}[\Mcal,C_p])\ev{w[C_p]},
\label{eq:wt-a}    
\end{equation}
\end{subequations}
where $\theta\in[0,2\pi)$, 
$\Mcal$ is a smooth, orientable, and closed $(D-p-1)$-dimensional submanifold of spacetime,
$\mathcal{B}$ is a $(D-p)$-dimensional manifold whose boundary is $\Mcal$,
and $\mathrm{Link}[\Mcal,C_p]$ denotes the linking number of $\Mcal$ and the $p$-dimensional cycle $C_p$.
Equations~\eqref{eq:wt-r} and~\eqref{eq:wt-a} show that the operators
\begin{equation}
    \exp(-\im\theta\int_\Mcal\star j),
    \qquad
     \exp(-\im\theta\int_\Mcal\star J)
\end{equation}
are the symmetry generators of 
$\rm U(1)_{\it r}^{[\it p]}$ and $\rm U(1)_{\it a}^{[\it p]}$, respectively.

\subsection{Construction of effective Lagrangian for \texorpdfstring{$p=1$}{p=1} and \texorpdfstring{$D=3+1$}{D=3+1}}

In the previous section, we identified the Nambu--Goldstone fields $A^{(p)}$ and $a^{(p)}$ associated with the symmetry breaking pattern ${\rm U}(1)_+^{[p]} \times {\rm U}(1)_-^{[p]} \to 1$. 
Here we construct the Schwinger-Keldysh effective Lagrangian 
in the case of $1$-form symmetry, $p=1$.

Our basic assumptions are as follows. 
We introduce no additional degrees of freedom beyond the doubled gauge fields. 
Since we consider a thermal system, Lorentz symmetry is explicitly broken, and the preferred rest frame is specified by a timelike four-vector $u^\mu$. 
For simplicity, we impose spacetime-translational and spatial rotational symmetries.\footnote{
Although condensed-matter systems often involve nontrivial interactions arising from broken rotational symmetry, ignoring such anisotropic effects does not alter the essential physics.}
Finally, we impose the dynamical KMS symmetry, which encodes the fact that the system is initially prepared in thermal equilibrium.

For $p=1$, the Nambu--Goldstone fields $A$ and $a$ are $1$-form gauge fields. 
The effective Lagrangian can therefore be constructed from the following:
\begin{equation}
    a_\mu,\ A_\mu,\ \p_\mu,\ \eta_{\mu\nu},\ \varepsilon_{\mu\nu\cdots},\ u^\mu,
\end{equation}
where $\eta_{\mu\nu}$ is the Minkowski metric and $\varepsilon^{\mu\nu\cdots}$ is the totally antisymmetric tensor. These are invariant tensors under Lorentz transformations.
For simplicity, we take
\begin{equation}
    u^\mu = (1,\overbrace{0,\cdots,0}^{D-1}).
\end{equation}
In $D=3+1$, it is convenient to use the following objects:
\begin{equation}
    (\bm{a}, \bm{A},)\
    \bm{e},\
    \bm{b},\
    \bm{E},\
    \bm{B},\
    \p_t,\
    \bm{\nabla} ,
    \label{GaugeInvBB}
\end{equation}
where $\bm e$ and $\bm b$ are the electric and magnetic fields associated with $a_\mu$, 
and $\bm E$ and $\bm B$ are those associated with $A_\mu$:
\begin{subequations}\begin{align}
    \bm{e} &\coloneqq -\dot{\bm a} -\bm\nabla a_0,
    &\bm{b} &\coloneqq \bm\nabla \times \bm a,\\
    \bm{E} &\coloneqq -\dot{\bm A} -\bm\nabla A_0,\
    &\bm{B} &\coloneqq \bm\nabla \times \bm A.
\end{align}\end{subequations}
With these, we construct a Lagrangian that is invariant under the $1$-form symmetry transformation, spacetime translation and spatial rotations up to total derivatives. 
Gauge-invariant terms can be expressed in terms of the electric and magnetic fields, which are themselves gauge invariant. 
In contrast, terms that are invariant only up to total derivatives cannot be written solely in terms of these fields and must involve the gauge potentials $\bm{a}$ and $\bm{A}$ explicitly.

\paragraph{Power counting.}
On constructing a low-energy EFT, we employ a derivative expansion, for which an appropriate power-counting scheme is required. Especially on a closed-time-path, it is important to adopt a power counting consistent with the dynamical KMS symmetry. As we note in appendix~\ref{app:DKMS-Calc}, the general form of the transformation rules on $a,A$ are given by
\begin{equation}
    \left\{\begin{aligned}
        \tilde A_0(-t,\eta \bm x)
        &= \eta_A A_0(t,\bm x),\\
        \tilde A_i(-t,\eta \bm x)
        &= -\eta_A\eta A_i(t,\bm x),\\
        \tilde a_0(-t,\eta \bm x)
        &= \eta_A (a_0(t,\bm x)
            + \im\beta \p_t A_0(t,\bm x)),\\
        \tilde a_i(-t,\eta \bm x)
        &= -\eta_A\eta (a_i(t,\bm x)
            + \im\beta \p_t A_i(t,\bm x)),
    \end{aligned}\right.
\end{equation}
where $\eta, \eta_A = \pm 1$.
We classify anti-Hermitian symmetries by the reversal properties of spatial coordinates and fields when gauge fields are vector fields; the resulting classification is given in table~\ref{tab:DefTheta}.

\begin{table}[tb]
\centering
\begin{tabular}{l|cccc}\hline
    $\Theta$&
    $\mathcal{T}$ &
    $\mathcal{PT}$ &
    $\mathcal{CT}$ &
    $\mathcal{CPT}$ \\ \hline
    $\eta$ & $+$ & $-$ & $+$ & $-$\\
    $\eta_A$ & $+$ & $+$ & $-$ & $-$\\\hline
\end{tabular}
\caption{The signs of $\eta$ and $\eta_A$ for different choices of $\Theta$. We assume the gauge fields are vector fields.
In the classification of \cite{Vardhan:2024qdi}, our choice corresponds to $\mathcal P_+$ and $\mathcal T_+$.
Different assignments of $\mathcal P$ and $\mathcal T$ simply amount to redefining the transformations via $\mathcal C$, namely $\mathcal P_- = \mathcal C\mathcal P_+$ and $\mathcal T_- = \mathcal C\mathcal T_+$; therefore, we restrict our discussion to this choice.}
\label{tab:DefTheta}
\end{table}

Since a dynamical KMS transformation mixes an $a$ field with a time derivative of an $r$ field, a consistent power counting satisfies $[\bm a] = [\bm A] + [\p_t]$.
As we will see below, the leading-order Lagrangian describes propagating electromagnetic waves; therefore, it is natural to assign the scaling $[\partial_t] = [\bm{\nabla}] = 1$. 
In addition, since an insulator is generally not expected to generate strong electromagnetic fields, we set $[\bm A] = 0$.
Then, the power counting for each of the building blocks in eq.~\eqref{GaugeInvBB} reads
\begin{align}
[\bm a]=1, \quad 
[\bm A]=0, \quad
[\bm e] = [\bm b] = 2, \quad
[\bm E] = [\bm B] = 1, \quad
[\p_t] = [\bm\nabla] = 1.
\end{align}

\paragraph{Invariant terms.}
Using the building blocks listed in eq.~\eqref{GaugeInvBB}, we can write down all terms consistent with rotational and internal symmetries. 
More generally, let $\mathcal{O}_{(m,n)}[a,A]$ denote an operator containing $m$ powers of $a$ and $n$ derivatives, then its power is counted as  $m + n$. 
Because of the unitarity condition \eqref{Unitarity1}, each term in the effective Lagrangian must include at least one power of $a$.
Below, we enumerate all candidate terms satisfying $m+n \le 4$:\footnote{
    As we will see later, this is the lowest order at which dissipative effects appear.
} \footnote{
    The terms $\bm{e}\cdot\dot{\bm B}$ and $\bm{b}\cdot\dot{\bm E}$ are equivalent up to total derivatives.
}
\begin{subequations}\begin{align}
    \Ocal_{(1,1)}[a,A]
    &= \bm{a}\cdot\bm{B},\\
    \Ocal_{(1,2)}[a,A]
    &=  \bm{e}\cdot\bm{E},\
        \bm{b}\cdot\bm{B},\
        \bm{b}\cdot\bm{E},\\
    \Ocal_{(2,1)}[a,A]
    &=  \bm{a}\cdot\bm{b},\\
    \Ocal_{(1,3)}[a,A]
    &=  \bm{e}\cdot\dot{\bm{E}},\
        \bm{b}\cdot\dot{\bm{B}},\
        \bm{e}\cdot\dot{\bm{B}},\
        \bm{b}\cdot\dot{\bm{E}},\
        \bm{b}\cdot(\bm\nabla\times\bm{B}),\;
        \bm{e}\cdot(\bm{E}\times\bm{B}),\
        \bm{b}\cdot(\bm{E}\times\bm{B}),\\
    \Ocal_{(2,2)}[a,A]
    &=  \bm{e}^2,\
        \bm{b}^2.
\end{align}
\label{SymmetricScalars}
\end{subequations}

We now further constrain these candidate terms listed by imposing the dynamical KMS symmetry, which encodes equilibrium constraints such as the fluctuation--dissipation relation in the Schwinger--Keldysh formalism.

At all orders of the power counting, the following terms
\begin{equation}
    \bm{a}\cdot\bm{B},\
    \bm{a}\cdot\bm{b},
    \label{CSterm4Dim}
\end{equation}
are the unique terms that are invariant under $\mathrm{U}(1)_+^{[1]}\times \mathrm{U}(1)_-^{[1]}$ symmetry up to total derivatives, yet cannot be expressed solely in terms of the gauge-invariant combinations ${\bm e, \bm b, \bm E, \bm B}$.
Such terms can be systematically constructed in arbitrary spacetime dimensions (see appendix~\ref{app:ChernSimonsForm}).
The terms in eq.~\eqref{CSterm4Dim} can form dynamical KMS symmetric terms by adding total derivatives as
\begin{equation}
    (\bm{a} - \im \beta \bm E)\cdot\bm{B},\
     (\bm{a}-\im\beta\bm{E})\cdot\bm{b}.
     \label{CSterm4Dim-kms}
\end{equation}
The first term in eq.~\eqref{CSterm4Dim-kms} encodes anomalous transport phenomena
called the chiral magnetic effect (CME). In appendix~\ref{sec:CME}, we give a detailed discussion on the physical effects of this term.
The latter term in eq.~\eqref{CSterm4Dim-kms} is, in fact, forbidden by the unitarity condition \eqref{Unitarity3}. As seen in eq.~\eqref{SymmetricScalars}, the $\bm{a}\cdot\bm{b}$ carries the lowest power $m+n=3$ among the $O(a^2)$ terms.
Therefore it dominates in the long-wavelength limit, with all others being suppressed.
Consequently, $\bm{a}\cdot\bm{b}$ would need to be positive-definite by itself.
However, this requirement is violated because the two helicity modes contribute with opposite signs.

The dynamical KMS–symmetric linear combinations consisting of the terms in eq.~\eqref{SymmetricScalars} are listed in table~\ref{tab:DKMSsymScalar} (see appendix~\ref{app:DKMS-symmetric-terms} for details).
None of the other combinations of terms in eq.~\eqref{SymmetricScalars} can form dynamical KMS–symmetric ones and are therefore excluded from the effective action in the near-equilibrium regime.\footnote{
In isotropic systems, the leading static magnetoelectric terms are
$\bm e \cdot \bm B$ and $\bm b \cdot \bm E$, both of which are total derivatives
and therefore do not produce a physical bulk response.
A nontrivial static magnetoelectric coupling can arise only in the presence of
additional low-energy degrees of freedom, such as a scalar field $\chi$,
through terms of the form $\chi \bm e \cdot \bm B$.
In the absence of such extra fields, isotropic systems do not admit a genuine
static magnetoelectric response.
By contrast, in anisotropic systems, one can construct electric–magnetic
cross-couplings that are not total derivatives, leading to a nontrivial
magnetoelectric response even without additional dynamical fields.
}

In table~\ref{tab:DKMSsymScalar}, `$+$' denotes that the term written in the left column satisfies dynamical KMS symmetry and thus permitted when $\Theta$ is taken as the top line.
`$-$' describes the term is basically forbidden but it is permitted together with $\Theta$-odd coefficient if the $\Theta$-symmetry is spontaneously broken
(see appendix~\ref{app:DKMS_SSB} for detail).
Transformations of each term under discrete symmetries without $\mathcal T$, namely $\mathcal C, \mathcal P$, and $\mathcal {CP}$, are also listed in the table.
In \cite{Vardhan:2024qdi}, it is shown that there are 11 different classes for the set of discrete symmetries, which leads to different effective field theories.
For example, if the set of discrete symmetries is $\{\mathcal T, \mathcal {PT}, \mathcal P\}$, we can choose $\Theta = \mathcal T$ or $\mathcal {PT}$ and additionally impose $\mathcal P$.
Up to the order $m+n\leq 4$, the 11 classes fall into only 2 categories; whether or not the system possesses a symmetry involving parity such as $\mathcal P, \mathcal {CP}, \mathcal {PT}$, or $\mathcal {CPT}$.
When it has such a parity-involved symmetry, the parity odd terms $\bm a \cdot \bm B$, $\bm e \cdot \dot{\bm B}$ and $\bm b \cdot \dot{\bm E}$ are excluded except when the symmetry is spontaneously broken.
If the system has no symmetry involving parity, then these terms are allowed.

\begin{table}[tb]
\centering
\begin{tabular}{l|cccc|ccc}\hline
    &
        $\mathcal{T}$ &
        $\mathcal{CT}$ &
        $\mathcal{PT}$ &
        $\mathcal{CPT}$ &
        $\mathcal{C}$ &
        $\mathcal{P}$ &
        $\mathcal{CP}$ \\ \hline
    $\bm a \cdot \bm B,\ 
     \bm e \cdot \dot{\bm B},\
     \bm b \cdot \dot{\bm E}$ &
        $+$ &
        $+$ &
        $-$ &
        $-$ &
        $+$ &
        $-$ &
        $-$  \\
    $\bm e \cdot \bm E,\
     \bm b \cdot \bm B,\ 
     \bm e \cdot (\bm e + \im\beta\dot{\bm E}),\
    \bm b \cdot (\bm b + \im\beta\dot{\bm B})$ &
        $+$ &
        $+$ &
        $+$ &
        $+$ &
        $+$ &
        $+$ &
        $+$ \\ \hline
\end{tabular}
\caption{
Summary of the conditions under which each term respects the dynamical KMS symmetry.
A `$+$' (`$-$') indicates that the term becomes dynamical KMS symmetric when accompanied by a $\Theta$-even (odd) coefficient.
Terms with $\Theta$-odd coefficients are allowed only when the $\Theta$ symmetry is spontaneously broken.
Transformations under discrete symmetries without $\mathcal T$, namely $\mathcal C, \mathcal P$, and $\mathcal C\mathcal P$, are also listed.}
\label{tab:DKMSsymScalar}
\end{table}

Let us examine the dynamical KMS transformations for several terms listed in table~\ref{tab:DKMSsymScalar}.
For instance, $\bm{e}\cdot\bm{E}$ is invariant under the dynamical KMS transformation up to total derivatives,
\begin{equation}\begin{split}
    \mathrm{T}_\Theta\cdot\mathrm{KMS}_\Theta
        [\bm{e}\cdot\bm{E}]
    - \bm{e}\cdot\bm{E}
    &= \p_t\qty(\frac{\im}{2}\beta\bm{E}^2).
    \label{DKMS_eE}
\end{split}\end{equation}
On the other hand, $\bm{e}\cdot(\bm{e}+\im\beta\dot{\bm E})$ and $ \bm{b}\cdot(\bm{b}+\im\beta\dot{\bm B})$ are strictly invariant under the dynamical KMS transformation, 
in the sense that no total-derivative terms are generated. This is due to
$\mathbb{Z}_2$ nature of $\Theta$-symmetry,
\begin{equation}\begin{split}
    &\mathrm{T}_\Theta\cdot\mathrm{KMS}_\Theta[
        \bm{e}\cdot(\bm{e}+\im\beta\dot{\bm E})
    ]
    - \bm{e}\cdot(\bm{e}+\im\beta\dot{\bm E})\\
    &= (\bm{e}+\im\beta\dot{\bm E})\cdot\bm{e}
        - \bm{e}\cdot(\bm{e}+\im\beta\dot{\bm E})\\
    &= 0.
    \label{DKMS_ee}
\end{split}\end{equation}

\paragraph{Effective Lagrangian.}
We thus obtain the most general effective Lagrangian
for electromagnetism in an insulating medium within the power-counting range $m+n \le 4$:\footnote{
Here, we do not include the term $\bm a \cdot \bm B$.
We discuss the physical effects associated with this term in appendix~\ref{sec:CME}.
}
\begin{equation}
    \Lcal_{\rm eff}[a,A]
    =\; \bm{e}\cdot\epsilon(
            \bm{E}
            - \tau_E\dot{\bm E}
            + v^2\gamma\dot{\bm B}
        )
        - \bm{b}\cdot\mu^{-1}\qty(
            \bm{B}
            + \tau_B\dot{\bm B}
            + \gamma \dot{\bm E}
        )
        + \frac{\im}{\beta}\qty(
            \tau_E\bm{e}\cdot\epsilon\bm{e}
            + \tau_B \bm{b}\cdot\mu^{-1}\bm{b}
        ),
    \label{Lagrangian}
\end{equation}
Here, $\epsilon, \tau_E, \gamma, \mu, \tau_B$ are transport coefficients
and we have defined $v \coloneqq 1/\sqrt{\epsilon\mu}$, which will be identified as the speed of light in the medium.
The coefficient $\gamma$ vanishes in the presence of an unbroken discrete symmetry involving parity.
The factors of $\mathrm{i}$ follow from the unitarity condition~\eqref{Unitarity2}.
The unitarity condition~\eqref{Unitarity3} further requires the transport coefficients to satisfy
\begin{equation}
    \epsilon\tau_{E} \ge 0, \quad \mu^{-1}\tau_B \ge 0.
\end{equation}

The dynamical KMS symmetry is verified by the fact that the effective Lagrangian changes only by a total derivative under the dynamical KMS transformation,\footnote{The term $\dot{\bm E} \cdot \dot{\bm B}$ is a total derivative, since 
$\dot{\bm E} \cdot \dot{\bm B} \propto \rd \dot{A} \wedge \rd \dot A = \rd (\dot A \wedge \rd \dot{A})$.
} 
\begin{equation}\begin{aligned}
    \mathrm{T}_\Theta\cdot\mathrm{KMS}_\Theta
        \{\Lcal_{\rm eff}[a,A]\}
    - \Lcal_{\rm eff}[a,A]
    &= \p_t\qty(
            \frac{\im}{2}\beta\bm{E}\cdot\epsilon\bm{E}
            - \frac{\im}{2}\beta\bm{B}\cdot\mu^{-1}\bm{B}
        ).
\end{aligned}\end{equation}
At this order, the resulting total derivative is independent of $a$, and no spatial total-derivative terms appear. 
This property does not necessarily persist once higher-derivative corrections are included.

The effective action contains the terms $\bm e\cdot\bm E$ and $\bm b\cdot\bm B$, which represent the most elementary contributions already present in the Maxwell theory.
As will be discussed in section~\ref{sec:Langevin}, terms $\bm e \cdot (\bm e + \im\beta\dot{\bm E})$ and $\bm b \cdot (\bm b + \im\beta\dot{\bm B})$ describe dielectric and magnetic losses, respectively. The parity-violating term $\bm b \cdot \dot{\bm E}$ represents a dynamical linear magnetoelectric effect; in particular, it accounts for the gyration of the light in parity-violating media.

We next comment on the terms $\bm e \cdot \dot{\bm B}$ and $\bm b \cdot \dot{\bm E}$.
These two terms are in fact interchangeable via integration by parts.
Each of them is separately invariant under the dynamical KMS symmetry.
In the Lagrangian~\eqref{Lagrangian}, we choose the specific combination
$\mu^{-1}\gamma (\bm e \cdot \dot{\bm B} - \bm b \cdot \dot{\bm E})$,
so that the resulting entropy current takes a simpler form, as discussed in
section~\ref{sec:entropy}.

Finally, let us comment on the relation to previous works.
Ref.~\cite{Vardhan:2024qdi} studied Schwinger--Keldysh effective field theories for a $1$-form symmetry, focusing primarily on the unbroken phase. The dissipative term appearing in eq.~\eqref{Lagrangian} is absent in their construction \changed{for the broken phase}.
Ref.~\cite{Salcedo:2024nex} also investigated Schwinger--Keldysh effective field theories for photons in the linear regime. Their Lagrangian, however, does not generically respect the $1$-form symmetry and includes gapped excitations. The dissipative terms in eq.~\eqref{Lagrangian} have not appeared explicitly in their formulation.
\changed{%
Ref.~\cite{christodoulidis} also constructed an EFT for the dynamics of electromagnetic fields within the Schwinger--Keldysh formalism.
They considered conducting media with Ohmic dissipation, in which the 1-form symmetry is explicitly broken, whereas in insulating media the symmetry is exactly preserved.}%
A recent work~\cite{Kaplanek:2025moq} analyzed Schwinger--Keldysh effective theories for photons from a top-down perspective, deriving the corresponding influence functional from a microscopic description. In particular, it was shown that the doubled BRST structure arises even if only a diagonal BRST symmetry is initially imposed.
\changed{%
Another distinctive features of our paper is that} we also provide a detailed discussion of the entropy current
for dissipative photons in section~\ref{sec:entropy}
and of electromagnetic duality in section~\ref{sec:duality},
which are not addressed in these works.

\section{Physical properties of the dissipative photon EFT}\label{sec:physical-properties}

In this section, we discuss the physical properties of the dissipative photon effective field theory constructed above. 
At finite temperature, the Schwinger--Keldysh effective action naturally encodes both the conservative and dissipative dynamics of the system, along with the corresponding stochastic fluctuations. 
We show that the effective theory can be reformulated as a set of Langevin-type equations describing the stochastic evolution of the gauge fields, from which the dispersion relations for the hydrodynamic photon modes can be derived (sections~\ref{sec:Langevin} and \ref{sec:dispersion}). 
We also discuss the Onsager reciprocal relations and the fluctuation-dissipation relation as consequences of the dynamical KMS symmetry (section~\ref{sec:Onsager&FDR}).
Furthermore, we identify the entropy current associated with the dissipative dynamics and confirm that its divergence is non-negative, ensuring consistency with the second law of thermodynamics (section~\ref{sec:entropy}).
Finally, we examine the electromagnetic duality of the dissipative photon EFT (section~\ref{sec:duality}).

\subsection{Equivalent Langevin description and physical interpretation}\label{sec:Langevin}

To clarify the physical content of the obtained effective theory, let us express it in terms of Langevin equations.
When the Schwinger-Keldysh effective action contains no cubic or higher-order terms in the $a$-type fields, it can be recast into a set of Langevin equations with Gaussian fluctuations. 
In the case of eq.~\eqref{Lagrangian}, these Langevin equations are obtained by performing a Hubbard--Stratonovich transformation, which introduces the auxiliary fields $\bm{\xi}_D$ and $\bm{\xi}_H$:
\begin{equation}\begin{split}
    &\int\Dcal a \Dcal A\
        \exp{\im\int\Lcal_{\rm eff}[a,A]}\\
    &\propto \int\Dcal a \Dcal A \Dcal\xi_{D,H}\
        \exp{\im\int\qty[
            \Lcal_{\rm eff}[a,A]
            + \frac{\im\beta}{4\epsilon\tau_E}\qty(
                \bm\xi_D
                - 2\im\frac{\tau_E}{\beta}\epsilon\bm{e}
              )^2
            + \frac{\im\beta\mu}{4\tau_B}\qty(
                \bm\xi_H
                - 2\im\frac{\tau_B}{\beta}\mu^{-1}\bm{b}
              )^2
        ]}\\
    &= \int\Dcal a \Dcal A \Dcal\xi_{D,H}\
        \exp{\im\int\qty[
            \bm{e}\cdot\bm{D}
            - \bm{b}\cdot\bm{H}
            + \frac{\im\beta}{4}(
                \epsilon^{-1}\tau_E^{-1}\bm\xi_D^2
                + \mu\tau_B^{-1}\bm\xi_H^2
              )
        ]},
        \label{Leff_HStrsf}
\end{split}\end{equation}
where we have defined $\bm{D}$ and $\bm{H}$ as
\begin{subequations}
\begin{align}
    \bm{D}
    &\coloneqq
        \epsilon (
            \bm{E}
            - \tau_E \dot{\bm E}
            + v^2\gamma\dot{\bm B}
        )
        + \bm{\xi}_D,\\
    \bm{H}
    &\coloneqq 
        \mu^{-1} (
            \bm{B}
            + \tau_B\dot{\bm B}
            + \gamma\dot{\bm E}
        )
        + \bm{\xi}_H.
\end{align}
\label{DefDandH}
\end{subequations}
In the last line of \eqref{Leff_HStrsf}, the Lagrangian is linear in the $a$-type fields. 
Performing the path integral over $a^\mu$ yields a delta functional that enforces 
the following equations of motion for $\bm D$ and $\bm H$,
\begin{subequations}\begin{align}
    \bm\nabla\cdot\bm{D} = 0,
    \label{GaussLaw}\\
    \dot{\bm{D}} - \bm\nabla\times\bm{H} = \bm0
    \label{AmpereLaw},
\end{align}\end{subequations}
where $\bm \xi_D$ and $\bm \xi_H$ are Gaussian white noise 
whose variances are given by
\begin{subequations}
\begin{align}
    \ev{\xi_{D}^i(x)\xi_D^j(y)}
    = \frac{2\epsilon\tau_E}{\beta}\delta^{ij}\delta^{(d)}(x-y),\\
    \ev{\xi_{H}^i(x)\xi_H^j(y)}
    = \frac{2\tau_B}{\beta\mu}\delta^{ij}\delta^{(d)}(x-y).
\end{align}\label{NoiseStrength}
\end{subequations}

It is now evident that $\boldsymbol{D}$ corresponds to the electric flux density and $\boldsymbol{H}$ to the magnetic field, as in the classical form of Maxwell’s equations in a medium. 
There is no contribution from an electric current in eq.~\eqref{AmpereLaw}, which confirms that the effective theory indeed describes an insulating system.
The remaining two equations follow from the Bianchi identity $\mathrm{d}F = 0$, which can be expressed in non-relativistic notation as
\begin{subequations}\begin{align}
    \boldsymbol{\nabla} \cdot \boldsymbol{B} &= 0,
    \label{MagGaussLaw}\\
    \dot{\boldsymbol{B}} + \boldsymbol{\nabla} \times \boldsymbol{E} &= \boldsymbol{0}.
    \label{FaradayLaw}
\end{align}\end{subequations}
These equations do not contain noise and hold even off-shell, in contrast to eqs.~\eqref{GaussLaw} and~\eqref{AmpereLaw}.

The physical meaning of the transport coefficients $\tau_E$ and $\tau_B$
is clear from eqs.~\eqref{DefDandH}.  
The coefficient $\tau_E$ represents the relaxation time \cite{BakerJarvis2007} with which the
polarization of the medium responds to variations of the electric field,
while $\tau_B$ characterizes the relaxation of the magnetization in response
to changes in the magnetic field.  
Both parameters therefore quantify the dissipative part of the electromagnetic
response of the medium.
A general relationship among the dynamical KMS symmetry, fluctuation-dissipation relations and Onsager's reciprocal relations will be presented in section~\ref{sec:Onsager&FDR}.

\subsection{Dispersion relation}
\label{sec:dispersion}
Let us now discuss the dispersion relation of photons within the dissipative
effective field theory.  Neglecting the noise terms, the equation of motion for 
the electric field follows from eqs.~\eqref{GaussLaw}, \eqref{AmpereLaw}, \eqref{MagGaussLaw}, and \eqref{FaradayLaw}, and it can be written as
\begin{align}
    \epsilon(
        \ddot{\bm{E}}
        - \tau_E\dddot{\bm{E}}
        - v^2\gamma\bm\nabla\times\ddot{\bm E}
    )
    + \bm\nabla\times\mu^{-1}[
        \bm\nabla\times(\bm{E}+\tau_B\dot{\bm{E}})
        - \gamma\ddot{\bm E}
    ]
    = \bm 0.
    \label{EOM-E}
\end{align}
To discuss linear excitations, we perform the Fourier transformation, $\bm E(t,\bm x) \sim \bm E_{\omega,\bm k} e^{-\im \omega t + \im \bm k \cdot \bm x}$. The equations of motion \eqref{EOM-E} can be diagonalized by 
taking helicity eigenstates,
$\im {\bm k} \times \bm E^{(h)}_{\omega,\bm k} = \lambda_h k \bm E^{(h)}_{\omega,\bm k}$, where $\lambda_h = \pm 1$ is the helicity eigenvalue and $k \coloneqq |\bm k|$.
For nontrivial solutions of eq.~\eqref{EOM-E} to exist, 
$\omega$ and $k$ must satisfy
\begin{equation}
    \epsilon(
        -\omega^2
        - \im \tau_E \omega^3
        + \lambda_h v^2 \gamma \omega^2 k
    )
    +  \mu^{-1} k (
        k
        - \im  \tau_B \omega k
        + \lambda_h \gamma \omega^2
      )
    = 0.
    \label{dispersion-eq}
\end{equation}
This equation gives the dispersion relation of the propagating electromagnetic mode.
Expanding the solution of eq.~\eqref{dispersion-eq} perturbatively in $k$, we obtain the low-momentum dispersion relation
\begin{equation}
    \omega(k)
    = \pm vk
        + \qty(
        \pm\lambda_h \gamma v
             - \im\frac{\tau_E+\tau_B}{2}
          )v^2 k^2
        + O(k^3),
\label{dispersion-pert}        
\end{equation}
where the signs are correlated. 
Equation~\eqref{dispersion-pert} includes both the propagating part and the dissipative correction induced by the transport coefficients $\tau_E$ and $\tau_B$.
For the vacuum to be stable, all excitation modes must be damped rather than
growing in time.  Combined with the unitarity constraint
\eqref{Unitarity3}, this requirement leads to the following stability
conditions:\footnote{
The conditions $\epsilon, \mu > 0$ follow from the requirement that the phase
velocity $v$ be real.  If either of them were negative, the dispersion relation
would give $\omega = \pm \im \Re(v)$, leading to exponentially growing,
and hence unstable, excitations.}
\begin{equation}
    \epsilon > 0,\
    \mu > 0,\
    \tau_E\ge0,\
    \tau_B\ge0.
\end{equation}

\subsection{Onsager reciprocity and fluctuation-dissipation relation}\label{sec:Onsager&FDR}

In the Schwinger--Keldysh effective theory, the dynamical KMS symmetry plays
a central role in constraining allowed terms in the effective Lagrangian.
In this subsection, we show that imposing the dynamical KMS symmetry
simultaneously leads to Onsager’s reciprocal relations and the
fluctuation-dissipation relations.

The dynamical KMS symmetry requires that the effective Lagrangian be invariant
up to total derivatives,
\begin{equation}
    \mathrm{T}_\Theta\cdot\mathrm{KMS}_\Theta[\Lcal_{\rm eff}]
    - \Lcal_{\rm eff}
    =\p_\mu V^\mu[a,A].
\end{equation}
It is convenient to decompose both the Lagrangian and the total derivative
term into their $\Theta$-even and $\Theta$-odd parts,
\begin{subequations}\begin{align}
    \Lcal_{\rm e}
    & \coloneqq \frac{1}{2}\qty(\Lcal_{\rm eff}+\mathrm{T}_\Theta\cdot\Theta[\Lcal_{\rm eff}]),
    &\Lcal_{\rm o}
    & \coloneqq \frac{1}{2}\qty(\Lcal_{\rm eff}-\mathrm{T}_\Theta\cdot\Theta[\Lcal_{\rm eff}]),\\
    V^\mu_{\rm e}
    & \coloneqq \frac{1}{2}\qty(V^\mu+\eta\ \mathrm{T}_\Theta\cdot\Theta [V^\mu]),
    &V^\mu_{\rm o}
    & \coloneqq \frac{1}{2}\qty(V^\mu-\eta\ \mathrm{T}_\Theta\cdot\Theta[V^\mu]),
\end{align}\end{subequations}
where the subscript $\rm e\ (\rm o)$ refers to $\Theta$-even (-odd) parts. 
The dynamical KMS condition then splits into two independent conditions,
\begin{subequations}\begin{align}
    \mathrm{T}_\Theta\cdot\mathrm{KMS}_\Theta[\Lcal_{\rm e}]
    - \Lcal_{\rm e}
    = \p_\mu V^\mu_{\rm e}[a,A],\\
    \mathrm{T}_\Theta\cdot\mathrm{KMS}_\Theta[\Lcal_{\rm o}]
    - \Lcal_{\rm o}
    = \p_\mu V^\mu_{\rm o}[a,A].
\end{align}\end{subequations}

The first condition constrains the $\Theta$-even (reactive) sector of the
effective theory and is the field-theoretic manifestation of
Onsager’s reciprocal relations.
The second condition constrains the $\Theta$-odd (dissipative) sector and
encodes the fluctuation-dissipation relations.

\paragraph{Examples of Onsager reciprocity.}

We first illustrate the Onsager reciprocal relations in the $\Theta$-even
sector. Consider 
\begin{equation}
        \Ocal_1[a,A]
        \coloneqq 
            \bm{e}\cdot(\dot{\bm{B}}\times\bm{B})
          + \bm{b}\cdot(\dot{\bm E} \times \bm{B}).    
\end{equation}
This combination is $\rm U(1)^{[1]}_+\times\rm U(1)^{[1]}_-$ symmetric, $\Theta$-even for $\Theta=\mathcal{T}$ or $\mathcal{PT}$, and then satisfies the dynamical KMS condition.
The dynamical KMS symmetry therefore enforces that 
the coefficients of the terms 
$\bm{e}\cdot(\dot{\bm{B}}\times\bm{B})$
and 
$\bm{b}\cdot(\dot{\bm E} \times \bm{B})$
be equal.
This reciprocity is a nonlinear extension of Onsager’s relations,
applicable beyond the linear-response regime.

As another example without time derivatives, consider
\begin{equation}\begin{split}
    \Ocal_2[a,A]
    &\coloneqq (\bm{e}\cdot\bm{E})\bm{B}^2
        + (\bm{b}\cdot\bm{B})\bm{E}^2.
\end{split}\end{equation}
This combination is again $\rm U(1)^{[1]}_+\times\rm U(1)^{[1]}_-$ symmetric, always $\Theta$-even, and dynamical KMS symmetric. 
Indeed, the dynamical KMS transformation of $\Ocal_2$ reads
\begin{equation}
    \mathrm{T}_\Theta\cdot\mathrm{KMS}_\Theta\{\Ocal_2[a,A]\}
    = \Ocal_2[a,A]
        + \frac{1}{2}\im\beta\partial_t[\bm{E}^2\bm{B}^2].
\end{equation}
This example again demonstrates that dynamical KMS symmetry enforces reciprocal couplings between electric and magnetic sectors.

\paragraph{Example of fluctuation-dissipation relation.}
Finally, we consider an example from the $\Theta$-odd sector,
\begin{equation}\begin{split}
    \Ocal_3[a,A]
    &\coloneqq \bm{e}\cdot(\bm{e}+\im\beta\dot{\bm E}).
\end{split}\end{equation}
This combination is $\rm U(1)^{[1]}_+\times\rm U(1)^{[1]}_-$ symmetric, always $\Theta$-even and dynamical KMS symmetric (see eq.~\eqref{DKMS_ee}). 
Here the first term represents fluctuations (noise), while the second term describes dissipation. Their fixed relative coefficient is precisely the fluctuation-dissipation relation.

\subsection{Entropy current}\label{sec:entropy}

In the framework of Schwinger--Keldysh effective field theories, the unitarity conditions together with the dynamical KMS symmetry impose strong constraints on the allowed dissipative structures. It has been shown that these principles make it possible to construct a local current $s^\mu$ whose divergence satisfies
\begin{equation}
    \p_\mu s^\mu \geq 0,
\end{equation}
ensuring consistency with the second law of thermodynamics.
The current $s^\mu$ is interpreted as the entropy current, characterizing
local entropy production associated with irreversible processes.

In this subsection, we explain how such an entropy current can be
systematically constructed for $p=1$-form SSB.
Our formulation clarifies how unitarity and dynamical KMS symmetry organize the structure of the entropy current.

For gauge invariance of the entropy, it is important to choose total derivatives for the Lagrangian to be a gauge invariant. Up to $a^2$ order, the effective Lagrangian is always written in following form:
\begin{equation}\begin{split}
    \Lcal_{\rm eff}[a,A]
    &= -\frac{1}{2}f_{\mu\nu}J_r^{\mu\nu}[A]
        + \im\Im(\Lcal_{\rm eff})[f,F],
\end{split}\end{equation}
where $J_{r}^{\mu\nu}$ is composed of $A^\mu(x)$ and its derivatives, and  is interpreted as the conserved current for the ${\rm U}(1)_r$ symmetry at $O(a^0)$,
\begin{equation}
    J_r[A]=J[a=0,A].
\end{equation}
By integrating by parts so that derivatives act only on the $a$-type fields,
the effective Lagrangian can be rewritten as
\begin{equation}\begin{split}
    \Lcal_{\rm eff}[a,A]
    &=  - a_\mu \p_\nu J_r^{\mu\nu}[A]
        + \frac{\im}{\beta}a_\mu\overrightarrow{\Sigma}^{\mu\nu}[A;\p]a_\nu
        + \p_\mu \mathcal{K}^\mu[a,A],
    \label{GaugeVarLcal}
\end{split}\end{equation}
where $\overrightarrow{\Sigma}^{\mu\nu}$ depends on $A^\mu(x)$ and its
derivatives, and the differential operator acts only on the field to its
right.

The dynamical KMS symmetry requires that the effective Lagrangian changes
only by a total derivative,
\begin{equation}
    \mathrm{T}_\Theta\cdot\mathrm{KMS}_\Theta\{\Lcal_{\rm eff}[a,A]\}
    - \Lcal_{\rm eff}[a,A]
    \eqqcolon
    \p_\mu (V^\mu[a,A]
    + \eta\ \mathrm{T}_\Theta\cdot\mathrm{KMS}_\Theta\{\mathcal{K}^\mu[a,A]\}
    - \mathcal{K}^\mu[a,A]) ,
\end{equation}
where $V^\mu$ denotes the contribution from the first two terms
in eq.~\eqref{GaugeVarLcal}.
As we will see below, the entropy current can be defined from 
$V^\mu$ and $\mathcal{K}^\mu$.

To find more explicit expression for $V^\mu$, let us check how each term in $\Lcal_{\rm eff}$ is transformed under the dynamical KMS transformation.
Since $\Theta$ is a $\mathbb{Z}_2$ transformation,
the current $J_r^{\mu\nu}$ can be decomposed into $\Theta$-even and $\Theta$-odd parts, 
\begin{subequations}\begin{align}
    J_{r\rm e}^{\mu\nu}[A]
    & \coloneqq \frac{1}{2}\qty(
        J_r^{\mu\nu}[A]
        + \eta_A\eta\ \mathrm{T}_\Theta\cdot\Theta\{J_r^{\mu\nu}[A]\}
    ),\\
    J_{r\rm o}^{\mu\nu}[A]
    & \coloneqq \frac{1}{2}\qty(
        J_r^{\mu\nu}[A]
        - \eta_A\eta\ \mathrm{T}_\Theta\cdot\Theta\{J_r^{\mu\nu}[A]\}
    ).
\end{align}\end{subequations}
In contrast, the second term in eq.~\eqref{GaugeVarLcal} is always $\Theta$-even, as it represents the highest-order term in $a$ in a dynamical-KMS-invariant combination; the justification is given in appendix~\ref{app:DKMS-Calc}.
Now we can perform the dynamical KMS transformation on these terms more explicitly,
\begin{subequations}\begin{align}
    \mathrm{T}_\Theta\cdot\mathrm{KMS}_\Theta
        \{a_\mu\p_\nu J_r^{\mu\nu}[A]\}
    &= (a_\mu+\im\beta\dot A_\mu)\p_\nu
        (J_{r\rm e}^{\mu\nu}[A] - J_{r\rm o}^{\mu\nu}[A]),\\
    \mathrm{T}_\Theta\cdot\mathrm{KMS}_\Theta
        \{a_\mu\overrightarrow{\Sigma}^{\mu\nu}[A;\p]a_\nu\}
    &= (a_\mu+\im\beta\dot A_\mu)\overrightarrow{\Sigma}^{\mu\nu}[A;\p](a_\nu+\im\beta\dot A_\nu).
\end{align}\end{subequations}
From these expressions, $V^\mu$ can be written in terms of
$J_r^{\mu\nu}$ and $\overrightarrow{\Sigma}^{\mu\nu}$ as
\begin{align}
    \p_\mu V^\mu[a,A]
    &= \im\beta\dot A_\mu\p_\nu J_{r\rm e}^{\mu\nu}[A]
        - (2a_\mu+\im\beta\dot A_\mu) \p_{\nu} J_{r\rm o}^{\mu\nu}[A] \nonumber \\
    & \quad    + (a_\mu+\im\beta\dot A_\mu)\overrightarrow{\Sigma}^{\mu\nu}[A;\p](a_\nu+\im\beta\dot A_\nu)
    -a_\mu\overrightarrow{\Sigma}^{\mu\nu}[A;\p]a_\nu.
\end{align}

The entropy current can now be written down, adopting the prescription developed in~\cite{Glorioso:2016gsa} to higher-form symmetries, as\footnote{Note that the expression depends on how to choose the total derivative terms.
Here, we have fixed them by requiring that each term in the Lagrangian is explicitly gauge-invariant.
}
\begin{equation}\begin{split}
    \p_\mu s^\mu[A]
    \coloneqq& [
            - \im \p_\mu V^\mu
            - \Im(\p_\mu \mathcal{K}^\mu)
        ]_{a=-\im\beta\dot{A}}\\
    =&  \beta \dot{A}_\mu\partial_\nu J_r^{\mu\nu}
        - [\Im(\Lcal_{\rm eff})]_{a=-\im\beta\dot{A}}\\
    =&  \beta \dot{A}_\mu\partial_\nu J_r^{\mu\nu}
        + [\Im(\Lcal_{\rm eff})]_{a=\beta\dot{A}}.
    \label{EntropyDeff}
\end{split}\end{equation}
In the last line, we used the fact that $\Im(\Lcal_{\rm eff})$ is $a^2$.
The first term in the last line is proportional,
\changed{%
up to $O(a^1)$,
to the equation of motion $0=\p_\nu J^{\mu\nu}= \p_\nu J_r^{\mu\nu}+O(a^1)$,
and therefore approximately vanishes in the expectation value, $\ev{\dot{A}_\mu\partial_\nu J_r^{\mu\nu}}\approx0$, as a consequence of the Ward--Takahashi identity for $\partial_{\nu}J^{\mu\nu}$ in the Schwinger-Keldysh contour%
}.
Because of the unitarity condition \eqref{Unitarity3}, 
the expectation value of the divergence of the entropy current is 
ensured to be locally positive semi-definite,
\changed{
\begin{equation}
    \ev{\partial_\mu s^\mu}
    =
\beta
\underbrace{
\ev{\dot{A}_\mu\partial_\nu J_r^{\mu\nu}}
}_{\approx 0}
      + \ev{\Im(\Lcal_{\rm eff})[f = \beta\dot{F},F]}
    \ge 0.
\end{equation}
}

In the case of the Lagrangian~\eqref{Lagrangian} we have constructed earlier, the entropy production rate is written as
\begin{equation}\begin{aligned}
    \p_\mu s^\mu
    &= \beta\qty[
            \dot A_0 \bm\nabla\cdot\overline{\bm{D}}
            - \dot{\bm A}\cdot(\dot{\overline{\bm{D}}}-\bm\nabla\times\overline{\bm{H}})
        ]
       +\beta(
            \tau_E\dot{\bm{E}}^2
            + \tau_B\dot{\bm{B}}^2
        )\\
    &\text{where}\qquad
    \left\{\begin{aligned}
        \overline{\bm{D}}
        &\coloneqq J_r^{0i}
        &&= \epsilon (\bm{E} - \tau_E\dot{\bm{E}} + v^2\gamma\dot{\bm B})\\
        \overline{\bm{H}}
        &\coloneqq \varepsilon_{0ijk} J_r^{jk}
        &&= \mu^{-1} (\bm{B} + \tau_B\dot{\bm{B}} + \gamma\dot{\bm{E}}).
    \end{aligned}\right.
\end{aligned}\end{equation}
The corresponding entropy current $s^\mu = (s^0, \bm s)$ can be identified as
\begin{equation}\begin{aligned}
    s^0
    &= \frac{1}{2}\beta\qty(
            \epsilon\bm{E}^2
            - \mu^{-1}\bm{B}^2
        )
       - \beta\bm{E}\cdot\overline{\bm{D}}
       + \beta A_0\bm\nabla\cdot\overline{\bm{D}},\\
    \bm{s}
    &= - \beta\bm{E}\times\overline{\bm{H}}
       - \beta A_0 \qty(
            \dot{\overline{\bm{D}}}
            -\bm\nabla\times\overline{\bm{H}}
         ).
    \label{EntropyCurr}
\end{aligned}\end{equation}
Off shell, the definition of the entropy current depends on the choice of gauge.
However, the gauge-dependent parts are proportional to the equations of motion and therefore vanish in expectation values or in correlations with physical quantities, namely, operators constructed solely from the $A$ fields, by virtue of the Ward--Takahashi identity.

\subsection{Electromagnetic duality}\label{sec:duality}

In this subsection, we examine how electromagnetic duality acts on the Schwinger--Keldysh effective theory derived above.
Although electromagnetic duality is usually formulated in a unitary setting, it is not obvious whether it extends to the dissipative effective theory.
We address this issue by implementing the duality directly as a transformation of the effective Lagrangian in the path-integral formulation.

Suppose we have an effective Lagrangian 
$\Lcal(\rd a, \rd A)$ in the broken phase, 
where $A^{(p)}$ and $a^{(p)}$ are $p$-form gauge fields
associated with ${\rm U}(1)_r^{[p]}$ and ${\rm U}(1)_a^{[p]}$ $p$-form symmetries, respectively.
We start with the path-integral representation of the generating functional,
\begin{equation}
    e^{W[b,B]} =
        \int\Dcal a \Dcal A
        \;\exp{\im\int\qty(
            \Lcal[\rd a, \rd A]
            + B\wedge\rd a
            + b\wedge\rd A
        )},
\end{equation}
where we introduced $(D-p-1)$-form background  gauge fields
$B$ and $b$ for magnetic ${\rm U}(1)_{r}^{[D-p-2]}$ and ${\rm U}(1)_{a}^{[D-p-2]}$ $(D-p-2)$-form symmetries, respectively.
Let us introduce $(p+1)$-form auxiliary fields $F$\footnote{
    Note that $\{f,F\}$ are defined as auxiliary fields in this section. It differs from the notation $f[a]=\rd a, \ F[A]=\rd A$ in other sections.
}
and a delta functional via 
\begin{equation}
    1
    = \int\Dcal F \;
        \delta(\rd A - F)
    = \int\Dcal F \Dcal\check{f}\;
        \exp{\im\int \check{f}\wedge(F-\rd A)},
\end{equation}
where $\check{f}$ is a $(D-p-1)$-form.
Using this delta-functional, $\rd A$ appearing 
in the exponent can be replaced with $F$.
Then, the integral over $A$ can be performed as 
\begin{equation}
    \int\Dcal A \;
        \exp{\im\int (-\check{f}\wedge \rd A)}
    = \delta(\rd\check{f}).
\end{equation}
This constrains that $\check{f}$ be a closed form.
By Poincaré's lemma, we can locally parameterize $\check{f}$ 
with a $(D-p-2)$-form gauge field 
$\check{a}$ as 
$\check{f} = \rd\check{a}$ and 
\begin{equation}
    \int\Dcal\check{f}\;
        \delta(\rd\check{f})\;
        \exp(\im\int \check{f}\wedge F)
    = \int\Dcal\check{a}\;
        \exp(\im\int \rd\check{a}\wedge F),
\end{equation}
with an appropriate choice of integral measure. Performing similar transformation about the $a$-field, we get
\begin{equation}
    e^{W[b,B]} 
    =  \int \Dcal f \Dcal F \Dcal\check{a} \Dcal\check{A}
        \;\exp{\im\int\qty[
            \Lcal[f, F]
            + (\rd\check{a} + b)\wedge F
            + (\rd\check{A} + B)\wedge f
        ]}.
    \label{DualBeforeIntF}
\end{equation}
When the Lagrangian is quadratic, 
the integration over $F$ and $f$ can be further performed,
and we obtain a dual Lagrangian,
\begin{equation}
    e^{W[b,B]} 
    =  \int \Dcal\check{a} \Dcal\check{A}
         \;\exp{\im\int
            \Lcal_{\rm dual} [\rd\check{a}+b, \rd\check{A}+B]
         }.
\end{equation}

Let us examine the duality  more explicitly for the Schwinger-Keldysh effective theory for
$p=1$ and $D=3+1$ derived earlier.
In this case, eq.~\eqref{DualBeforeIntF} involves the following functional integral,
\begin{equation}
    \int \Dcal\bm E
    \exp{\im\int
    \rd^4 x
    \qty[
        \bm e \cdot \epsilon(
            \bm E 
            - \tau_E \dot{\bm E}
        )
        - \bm b \cdot \mu^{-1}\gamma\dot{\bm E}
        + \varepsilon^{0ijk}(\p_i\check{a}_j+b_{ij}) E_k
    ]},
\end{equation}
where we have decomposed the integration measure as $\Dcal F = \Dcal\bm E\ \Dcal\bm B.$
The dual theory is obtained by performing the integrals $\Dcal f$ and $\Dcal F$. 
The integration of $F$ produces the delta functionals,
\begin{align}
    \delta\qty[
        \mqty(
          \mu^{-1}\gamma\p_t &
          \mu^{-1}(1-\tau_B\p_t)\\
          -\epsilon(1+\tau_E\p_t) &
          -v^2\epsilon\gamma\p_t
        )
        \mqty(
            \bm e\\
            \bm b
        )
        \changed{+}
        \mqty(
            -\bm\nabla\changed{\check a^0} -\p_t\check{\bm a}\changed{-(b_{0i})}\\
            \bm\nabla\times\check{\bm a}\changed{-(\varepsilon^{0ijk}b_{jk})}
        )
    ].
\end{align}
These delta functionals enforce the replacement
\begin{equation}
    (\bm E,\bm B)
    \longmapsto
    (\check{\bm E}, \check{\bm B})
    \simeq
    \changed{(-\bm H, \bm D)}.
\end{equation}
Since we focus on a sufficiently low-energy regime,
we perform a derivative expansion,
\begin{equation}
    (1 + \tau_E\p_t)^{-1}
    = 1 - \tau_E\p_t + \cdots,
\end{equation}
and drop higher-order terms.
Then the $\Dcal f = \Dcal \bm{e}\ \Dcal \bm{b}$ integral is performed using the delta functional. The resulting dual Lagrangian takes the form
\begin{equation}\begin{split}
    \Lcal_{\rm dual}[\rd\check{a}, \rd\check{A}]
    = \check{\bm e}\cdot\check\epsilon\qty(
            \check{\bm E}
            - \check\tau_E \dot{\check{\bm E}}
            + v^2\check\gamma\dot{\check{\bm B}}
        )
      - \check{\bm{b}}\cdot\check\mu^{-1}\qty(
            \check{\bm B}
            + \check\tau_B \dot{\check{\bm B}}
            + \check\gamma \dot{\check{\bm E}}
        )
      + \frac{\im}{\beta}\qty(
            \check\tau_E\check{\bm e}\cdot\check\epsilon\check{\bm e}
            + \check\tau_B\check{\bm b}\cdot\check\mu^{-1}\check{\bm b}
        )\\
    \text{where}\quad
    \check{\bm e}
    \coloneqq -\bm\nabla\check{a}_0 - \dot{\check{\bm a}},
    \quad
    \check{\bm E}
    \coloneqq -\bm\nabla\check{A}_0 - \dot{\check{\bm A}},
    \quad
    \check{\bm b}
    \coloneqq \bm{\nabla}\times\check{\bm a},
    \quad
    \check{\bm B}
    \coloneqq \bm{\nabla}\times {\check{\bm A}},
\end{split}\end{equation}
up to the order $m+n\le4$.
The dual transformation of effective theory parameters is
\begin{equation}
    \left\{\begin{aligned}
        \check\epsilon
        &= \mu,
        &\check\tau_E
        &= \tau_B,\\
        \check\mu
        &= \epsilon,
        &\check\tau_B
        &= \tau_E,
    \end{aligned}\right.
    \quad\check\gamma=\gamma.
\end{equation}
This demonstrates that the roles of $\bm E$ and $\bm B$ are interchanged in the dual theory, consistent with electromagnetic duality in the absence of dissipation.

\section{EFT for an unbroken \texorpdfstring{${\rm U}(1)$}{U(1)} 1-form symmetry}\label{sec:unbroken}

In the previous sections, we have mainly focused on the effective field theory in phases where the ${\rm U}(1)$ $1$-form symmetry is spontaneously broken. In such cases, the low-energy dynamics is governed by the corresponding Nambu--Goldstone modes. In this section, we turn to the complementary situation and briefly discuss the effective field theory for an \emph{unbroken} ${\rm U}(1)$ $1$-form symmetry. Although no propagating mode appears in this phase, the Schwinger--Keldysh formulation exhibits characteristic emergent redundancies, which play a crucial role in governing diffusive dynamics. Our aim here is to clarify the origin of these redundancies and to construct the corresponding effective theory in the unbroken phase.

\subsection{Origin of diffusive symmetry}\label{sec:origin-diffusive-symmetry}

For an unbroken ${\rm U}(1)$ 0-form symmetry, the Schwinger--Keldysh EFT exhibits an emergent gauge redundancy in the $r$-sector, namely a spatially dependent shift of the coset variable,
$\pi_r(t,\bm x)\mapsto \pi_r(t,\bm x)+\lambda(\bm x)$,
with $\lambda$ independent of time.
This \emph{diffusive symmetry} is the symmetry principle underlying charge diffusion in the unbroken phase \cite{Crossley:2015evo}.
Relatedly, a recent viewpoint interprets the diffusion mode as a Nambu-Goldstone mode of strong-to-weak spontaneous symmetry breaking~\cite{Lessa:2024gcw, Gu:2024wgc, Huang:2024rml}, providing a physical explanation for the unusual reparameterization/shift-type redundancies required in the
Schwinger--Keldysh EFT of diffusion \cite{Huang:2024rml} (seel also \cite{Akyuz:2023lsm}).

We now examine the origin of the diffusive symmetry in the case of an unbroken ${\rm U}(1)$ symmetry.
Let us begin with a ${\rm U}(1)$ $0$-form symmetry.
In the unbroken phase, the low-energy degrees of freedom are those associated
with the coset corresponding to
\changed{%
the symmetry-breaking pattern
\begin{equation}
{\rm U}(1)_r \times {\rm U}(1)_a
\to 
{\rm U}(1)_r.
\end{equation}
}%
Rather than parametrizing the coset directly, 
we posit that the symmetry is fully broken,
${\rm U}(1)_r \times {\rm U}(1)_a \to 1$,
and later mod out the additional degrees of freedom as a gauge redundancy.

Let us consider an effective action $I_{\rm eff}$ in the presence of a background gauge field $A_r$ coupled to the ${\rm U}(1)_r$ $0$-form symmetry.
We introduce a gauge-invariant combination
\begin{equation}
\mathcal A_r \coloneqq A_r + \rd \pi_r .
\end{equation}
The effective action can be written as a function of $\mathcal A_r$, 
$I_{\rm eff}[\mathcal A_r]$.\footnote{The action also depends on $a$-type fields, but here we suppress the dependence for notational simplicity.}
In equilibrium, where $\rd \pi_r = 0$, the system is assumed to be in a thermal state characterized by thermodynamic parameters such as the temperature and chemical potential.
The chemical potential is identified with the temporal component of the gauge
field, $(A_r)_0 = \mu$.\footnote{
More generally, one may introduce Lagrangian coordinates $(\sigma^0,\sigma^i)$ and consider an embedding $x^\mu = x^\mu(\sigma^0,\sigma^i)$. This formulation allows for a description of local thermal equilibrium with a spacetime-dependent temperature $T(x)$ and chemical potential $\mu(x)$. In the present work, however, we restrict ourselves to the case in which the $\sigma$ coordinates describe a flat Minkowski space and focus on global thermal equilibrium in order to simplify the notation. The extension to local thermal equilibrium can be implemented straightforwardly.
}
More generally, introducing a timelike vector field
$u = u^\mu \partial_\mu$ specifying the local rest frame,
the chemical potential may be written as
\begin{equation}
\mu = i_u A_r ,
\end{equation}
where $i_u$ denotes the interior product, $i_u A_r = u^\mu (A_r)_\mu$.

At first sight, one might expect the effective action to be invariant under the full gauge redundancy
\begin{equation}
I_{\rm eff}[\mathcal A_r] 
=
I_{\rm eff}[\mathcal A'_r] ,
\quad \mathcal A'_r = \mathcal A_r + \rd \lambda ,
\end{equation}
with a generic gauge parameter $\lambda$.
However, in hydrodynamic effective field theory the thermal equilibrium state
serves as the reference configuration about which the theory is constructed.
Accordingly, the thermodynamic parameters characterizing this state, in
particular the chemical potential $\mu$, are regarded as fixed physical parameters.\footnote{This viewpoint has been emphasized in \cite{Glorioso:2018kcp} for both 0-form and $1$-form symmetries.}

This implies that admissible gauge transformations must preserve the value of
the chemical potential,
\begin{equation}
i_u A_r = i_u A'_r .
\end{equation}
which leads to the constraint
\begin{equation}
i_u \rd \lambda = 0 .
\label{eq:gauge-cond}
\end{equation}
For a static equilibrium configuration with $u^\mu = (1,\bm 0)$, this condition reduces to
\begin{equation}
    \partial_t \lambda (t, \bm x) = 0,
\end{equation}
implying that allowed gauge transformations are time-independent, 
$\lambda = \lambda(\bm x)$.

As the effective action with fluctuation obtained by the replacement 
$I_{\rm eff}[A_r] \mapsto I_{\rm eff}[\mathcal A_r]$, $I_{\rm eff}[\mathcal A_r]$ should have the same gauge redundancy under $\mathcal A_r \mapsto \mathcal A_r + \rd \lambda(\bm x)$.
Since we mod out the gauge degrees of freedom in the unbroken phase, this gauge invariance must persist
even in the absence of background gauge fields $A_r$.
Accordingly, we require the action to be invariant under the following shift of the coset variable,
\begin{equation}
\pi_r (t, \bm x) \mapsto \pi_r (t, \bm x) + \lambda(\bm x).
\end{equation}
We thus conclude that this restricted gauge redundancy, which originates from
fixing the thermodynamic parameters of the equilibrium state, gives rise to the diffusive shift symmetry in the Schwinger--Keldysh effective field theory.

The above argument generalizes directly to higher-form symmetries.
For $1$-form symmetries, analogous restrictions on the gauge parameter arise from requiring the invariance of the string chemical potential, as discussed in \cite{Glorioso:2018kcp}. In what follows, we reinterpret this restricted gauge invariance as a residual gauge symmetry of the Schwinger--Keldysh coset variables in the unbroken phase.
We have a background 2-form gauge field $B_r$ for the ${\rm U}(1)_r$ $1$-form symmetry, and its gauge transformation is written as
\begin{equation}
    B_r \mapsto B'_r = B_r + \rd \Lambda. 
\end{equation}
We again restrict to gauge transformations preserving the thermodynamic potential:
\begin{equation}
i_u B_r = i_u B'_r .
\end{equation}
This leads to
\begin{equation}
i_u \rd \Lambda
= u^\mu (\p_\mu \Lambda_\nu - \p_\nu \Lambda_\mu) \rd x^\nu 
\propto (\p_t \Lambda_i - \p_i \Lambda_t) \rd x^i = 0.
\label{eq:i-alpha-d-lambda}
\end{equation}
Thus the allowed gauge parameters must be of the form\footnote{
A generic local solution for eq.~\eqref{eq:i-alpha-d-lambda} is given by $
\Lambda_\mu = \partial_\mu \chi(t,\bm x) + \widetilde{\Lambda}_\mu$, where $\widetilde{\Lambda}_\mu$ satisfies
\begin{equation}
\p_t \widetilde\Lambda_i = 0, 
\quad 
\p_i \widetilde\Lambda_t = 0.
\end{equation}
As $\p_\mu \chi$ can be absorbed into the redundancy of the gauge parameters for the $1$-form symmetry, we arrive at the parameters of the form~\eqref{eq:1-form-lambda}.
}
\begin{equation}
\Lambda_i = \Lambda_i (\bm x), 
\quad 
\Lambda_t = \Lambda_t (t).
\label{eq:1-form-lambda}
\end{equation}
Moreover, since $\Lambda_t(t)\rd t$ can be regarded as a $1$-form
on the $1$-dimensional time manifold, it is necessarily exact and can be written as $\Lambda_t(t)\rd t = \p_t \chi(t) \rd t$, and hence can be gauged away. We therefore conclude that, in the unbroken phase, the effective theory should be invariant under the transformation of the dynamical $1$-form gauge field
$A_r \mapsto A_r + \Lambda$, where $\Lambda = \Lambda_i(\bm x)\rd x^i$, \changed{in addition to $\mathrm{U}(1)_+^{[1]}\times \mathrm{U}(1)_-^{[1]}$ transformations}.

\changed{%
We may ask whether the resulting theory is consistent with the symmetry being unbroken in this phase.
This is indeed the case.
Once the diffusive symmetry is imposed, the Wilson loop
$W(C)={\rm exp}\big(\im\oint_C A_r \big)$ 
becomes gauge-variant.
As a result, its expectation value vanishes, $\langle W(C) \rangle = 0$, as expected for an unbroken symmetry.
}%

\subsection{Effective Lagrangian in \texorpdfstring{$D=3+1$}{D=3+1}}

For concreteness, let us consider $D = 3+1$
and construct the effective Lagrangian for an unbroken ${\rm U}(1)$ $1$-form symmetry.%
\footnote{\changed{%
Although the EFT for the unbroken phase is formulated as magnetohydrodynamics in \cite{Vardhan:2024qdi},
our analysis employs a different power-counting scheme (see eq.~\eqref{eq:generalcounting}).
}}
Under the restricted transformations \eqref{eq:1-form-lambda}, the electromagnetic fields,
$
\bm E =- \dot{\bm A} - \bm\nabla A_0,
\bm B = \bm \nabla \times \bm A,
$
transform as
\begin{equation}
\bm E' = \bm E, 
\quad 
\bm B' = \bm B + \bm \nabla \times \bm \Lambda (\bm x).
\label{eq:e-b-shift-unbroken}
\end{equation}
The electric field remains invariant, while the magnetic field shifts by a spatial curl.
Accordingly, the invariant objects that we can use to write down the Lagrangian are as follows:
\begin{equation}
    \bm{e},\
    \bm{b},\
    \bm{E},\
    \dot{\bm{B}},\
    \p_t,\
    \bm{\nabla} ,
    \label{GaugeInvBB_unbroken}
\end{equation}

\paragraph{Power counting.}
Let us fix the power counting.
The effective Lagrangian to the leading order reads 
\begin{equation}
    (\Lcal_{\rm eff})_0[a,A]
    = \bm{e}\cdot\epsilon \bm E
        - \bm{b}\cdot\mu^{-1} \tau_B\dot{\bm{B}}.
    \label{Lagrangian-unbroken-phase-leading}
\end{equation}
The leading-order Lagrangian indicates that we should have
\begin{equation}
e E \sim b \p_t B .
\end{equation}
The Bianchi identity implies 
$\p_t b \sim \nabla e$ and $\p_t B \sim \nabla E$, so 
\begin{equation}
e E \sim b \p_t B \sim \frac{\nabla}{\p_t} e\nabla E = \frac{\nabla^2}{\p_t}e E .
\end{equation}
This suggests $\p_t \sim \nabla^2$, and we shall set 
\begin{equation}
[\bm\nabla] = 1, \quad [\p_t] = 2.
\end{equation}
Combining this with the Bianchi identity, we should have
$E \sim \nabla B.$
Thus, the electric field is higher-order from the magnetic field by a spatial derivative,
\begin{equation}
[\bm E] =[\bm B] + 1, 
\quad 
[\bm e] =[\bm b] + 1.
\end{equation}
From the dynamical KMS symmetry, 
$[\bm e] = [\dot{\bm E}] = [\bm E] + 2$,
and
$[\bm b] = [\dot{\bm B}] = [\bm B] + 2$.
By setting $[\bm A] = \alpha$,
the power counting of electromagnetic fields is given by
\begin{align}
[\bm B]=\alpha + 1 ,\quad
[\bm E]=\alpha + 2 ,\quad
[\bm b]=\alpha + 3 ,\quad
[\bm e]=\alpha + 4.
\label{eq:generalcounting}
\end{align}
In \cite{Vardhan:2024qdi}, the strong-field regime is considered, which corresponds to the choice $\alpha = -2$.
Here, let us employ $\alpha = -1$, with which the electromagnetic fields are counted as 
\begin{equation}
[\bm B]=0, \quad 
[\bm E]=1, \quad
[\bm b]=2, \quad
[\bm e]=3.
\end{equation}

\begin{table}[tb]
\centering
\begin{tabular}{l|cccc|ccc}\hline
    &
        $\mathcal{T}$ &
        $\mathcal{CT}$ &
        $\mathcal{PT}$ &
        $\mathcal{CPT}$ &
        $\mathcal{C}$ &
        $\mathcal{P}$ &
        $\mathcal{CP}$ \\ 
        \hline
  $\bm e \cdot \bm E, \,\, \bm b \cdot (\bm b + \im \beta \dot{\bm B})$ &
        $+$ &
        $+$ &
        $+$ &
        $+$ &
        $+$ &
        $+$ &
        $+$ \\        
$\bm b \cdot \dot{\bm E}$    
    &
        $+$ &
        $+$ &
        $-$ &
        $-$ &
        $+$ &
        $-$ &
        $-$  \\
  $\bm b \cdot (\bm E \times \dot{\bm B})$ &
        $-$ &
        $+$ &
        $+$ &
        $-$ &
        $-$ &
        $-$ &
        $+$ \\
        \hline
\end{tabular}
\caption{
Summary of the conditions under which each term respects the dynamical KMS symmetry.
A `$+$' (`$-$') indicates that the term becomes dynamical KMS symmetric when accompanied by a $\Theta$-even (odd) coefficient.
Terms with $\Theta$-odd coefficients are allowed only when the $\Theta$ symmetry is spontaneously broken.
Transformations under discrete symmetries without $\mathcal T$, namely $\mathcal C, \mathcal P$, and $\mathcal C\mathcal P$, are also listed.}
\label{tab:DKMSsymScalar-unbroken}
\end{table}

\paragraph{Effective Lagrangian.}
Now, we are ready to list up possible terms order by order. We use the forms where the invariance under the dynamical KMS transformation is manifest. 
Leading-order terms are of order 4, and allowed terms are
\begin{equation}
\bm e \cdot \bm E,
\quad
\bm b \cdot (\bm b + \im \beta \dot{\bm B}).
\end{equation}
Note that these combinations are allowed for any choice of $\Theta$, as they are always even under $\Theta$.
At order 5,\footnote{
Note that $\bm e \cdot (\bm\nabla \times \bm E) \sim \bm b \cdot \dot{\bm E}$ up to a total derivative.
}
\begin{equation}
\bm b \cdot \dot{\bm E}, \quad 
\bm b \cdot (\bm E \times \dot{\bm B}), \quad 
\im (\bm b + \im \beta \dot{\bm B})\cdot (\bm \nabla \times \bm b).
\label{eq:unbroken-order-5}
\end{equation}
We note that the third combination in eq.~\eqref{eq:unbroken-order-5} is in fact not allowed as $\bm b \cdot \bm\nabla \times \bm b$ can have either sign, which is inconsistent with the positivity~\eqref{Unitarity3} of the imaginary part of the Lagrangian.
In table~\ref{tab:DKMSsymScalar-unbroken}, we list how the allowed terms transform under different choices of $\Theta$.
They can appear in the Lagrangian when they are $\Theta$-even.
These terms are also allowed when they are $\Theta$-odd and 
the corresponding $\Theta$ is spontaneously broken. 

Hence, a general effective Lagrangian in the unbroken phase up to the fifth order is\footnote{
The presence of the term $\bm b \cdot (\bm E \times \dot{\bm B})$, 
which can appear \changed{when the system does not respect any of $\mathcal{T}$, $\mathcal{CPT}$, $\mathcal{C}$, and $\mathcal{P}$}, corresponds to the difference between D1 and D2 in \cite{Vardhan:2024qdi}.
Note that our $\mathcal T$ and $\mathcal P$ correspond to $\mathcal{T_+=CT_-}$ and $\mathcal{P_+=CP_-}$, respectively, in the notation of \cite{Vardhan:2024qdi}.
}
\begin{equation}
    \Lcal_{\rm eff}[a,A]
    = \bm{e}\cdot\epsilon \qty(
        \bm{E}
        - v^2\gamma \bm\nabla \times \changed{\bm E}
      )
      - \bm{b}\cdot\mu^{-1}\qty(
          \tau_B\dot{\bm{B}}
          + \gamma \dot{\bm E}
          + \kappa \bm{E}\times \dot{\bm B}
        )
      + \frac{\im}{\beta} \tau_B 
        \bm{b}\cdot\mu^{-1}\bm{b}.
    \label{Lagrangian-unbroken-phase}
\end{equation}
The corresponding equations of motion written in terms of $\bm E$ are 
\begin{subequations}\begin{align}
\epsilon\qty(
  \dot{\bm E}
  - v^2\gamma \bm\nabla \times \changed{\dot{\bm E}}
)
- \mu^{-1} 
\left(
\tau_B  {\bm \nabla}^2 \bm E 
+ \gamma \bm \nabla \times \dot{\bm E}
+ \kappa \bm \nabla \times (\bm E \times(-\bm\nabla \times \bm E))
\right)
&= \bm 0, \\
\bm \nabla \cdot \bm{E}  &= 0.
\end{align}\end{subequations}
Let us find the dispersion relations for linear excitations.
Using helicity eigenstates, these equations can be diagonalized, 
and the dispersion relation of linear modes is obtained, up to $O(k^3)$, as
\begin{equation}
\omega (k) 
=
- \im \tau_B v^2 k^2 
- 2\im \lambda_h \gamma \tau_B v^4 k^3 
+ O(k^4) ,
\end{equation}
where $v\coloneqq 1/\sqrt{\epsilon \mu}$, and 
$\lambda_h = \pm 1$ denotes the helicity.
Unlike the case of a spontaneously broken ${\rm U}(1)$ symmetry, 
the modes are purely diffusive,
with a helicity-dependent correction appearing at $k^3$.

The entropy production in this phase is given by eq.~\eqref{EntropyDeff}, since the effective Lagrangian is at most quadratic in the $a$-type fields.
The resulting entropy current is
\begin{equation}\begin{aligned}
    s^0
    &= \frac{1}{2}\beta\epsilon\bm{E}^2
       - \beta\bm{E}\cdot\overline{\bm{D}}
       + \beta A_0\bm\nabla\cdot\overline{\bm{D}},\\
    \bm{s}
    &= - \beta\bm{E}\times\overline{\bm{H}}
       - \beta A_0 \qty(
            \dot{\overline{\bm{D}}}
            - \bm\nabla\times\overline{\bm{H}}
         ).
\end{aligned}\end{equation}
where $\overline{\bm D}$ and $\overline{\bm H}$ are given by
\begin{equation}
    \left\{\begin{aligned}
        \overline{\bm{D}}
        &\coloneqq J_r^{0i}
        &&= \epsilon\qty(
                \bm{E}
                + v^2\gamma\dot{\bm B}
            )            ,\\
        \overline{\bm H}
        &\coloneqq
            \varepsilon_{0ijk} J_r^{jk}
        &&= \mu^{-1} \qty(
                \tau_B\dot{\bm B}
                + \gamma\dot{\bm E}
                + \kappa\bm{E}\times\dot{\bm B}
            ).
    \end{aligned}\right.
\end{equation}

\changed{%
Finally, we briefly comment on the physical realization of our effective Lagrangian.
In terms of the conserved currents $\overline{\bm D}$ and $\overline{\bm H}$, the equations of motion in the absence of fluctuations are
\begin{align}
    &\dot{\overline{\bm D}} = \bm\nabla\times \overline{\bm H} 
    \simeq \left[
        \tau_B v^2\Delta \overline{\bm D} + 2\gamma \tau_B v^4 \bm\nabla\times\Delta\overline{\bm D}
        -\kappa \epsilon^{-1}v^2 \bm\nabla\times \left(\overline{\bm D}\times(\bm\nabla\times\overline{\bm D})\right)
    \right], \\
    &\bm\nabla \cdot \overline{\bm D} = 0,
\end{align}
which, at the leading order, describe the diffusion of the transverse conserved density $J_r^{0i} = \overline{\bm D}$.
When the $1$-form symmetry under consideration is the magnetic ${\rm U}(1)$ and remains unbroken, the conservation law is given by the Bianchi identity $\partial_{\mu}\tilde F^{\mu\nu}=0$, where $\tilde F^{\mu\nu} = \frac{1}{2}\epsilon^{\mu\nu\rho\sigma}F_{\rho\sigma}$, and the conserved current is $J_r^{\mu\nu} = \tilde F^{\mu\nu}$.
Thus, the above equation can be interpreted as a diffusion equation for the magnetic flux density $J_r^{0i} = -\overline{\bm B}$.\footnote{
In this particular example, the conserved current $\tilde F^{\mu\nu}$ does not coincide with the conventional definition of the field strength tensor in terms of the gauge field, and thus the two should not be confused.
}
This magnetic diffusion equation corresponds to one of the fundamental equations in magnetohydrodynamics, where a charged current is present and the electric $1$-form symmetry is explicitly broken.
}%

\section{Relation to strong/weak symmetries}\label{sec:strongweaksym}

In this section, we discuss the relation between the Schwinger--Keldysh formulation of effective field theories and the notions of strong and weak symmetries~\cite{Buča_2012,PhysRevA.89.022118,Lessa:2024gcw,Gu:2024wgc,Huang:2024rml}. The distinction between strong and weak symmetries was introduced and systematized in the context of open quantum systems, where the two types of symmetry act differently on density matrices.
A strong symmetry can impose nontrivial constraints on the quantum dynamics, such as the existence of invariant subspaces, degeneracies of stationary states, or the protection of quantum coherence against decoherence~\cite{Buča_2012,PhysRevA.89.022118,Akamatsu:2021vsh}.
In contrast, thermal or stationary states often preserve only a weaker, diagonal symmetry even when the underlying dynamics admits a strong symmetry. In the following, we examine these notions from the perspective of the Schwinger--Keldysh formalism.

Let us start by recalling the definition of weak and strong symmetries. 
For a symmetry group $G$, let each element $g \in G$ act on the Hilbert space through a unitary operator $U(g)$. 
The action of a weak symmetry on the density matrix is defined as
\begin{equation}
\mathcal U_{\rm w}(g) [\rho]
\coloneqq U(g) \rho U^\dag (g),
\end{equation}
whereas a strong symmetry corresponds to an independent action on either the ket or bra indices of the density matrix, namely,
\begin{equation}
\mathcal U_{\rm s}(g) [\rho] \coloneqq U(g) \rho 
\quad \text{or} \quad 
\mathcal U_{\rm s}(g) [\rho] \coloneqq \rho U^\dag (g).
\end{equation}
We consider an open quantum system whose dynamics is governed by a Lindblad equation, $\partial_t \rho = \mathcal L[\rho]$, with $\mathcal L$ the corresponding Liouvillian superoperator.
The system is said to respect a weak (strong) symmetry $G$ if 
its action commutes with the Liouvillian,
\begin{equation}
[\mathcal L , \mathcal U_{\rm w(s)}(g)] = 0,
\end{equation}
for any $g \in G$.

Let us now compare this structure with the Schwinger--Keldysh formulation (see \cite{Akyuz:2023lsm}). Suppose that the system has a symmetry $\mathcal G$ which is spontaneously broken as $\mathcal G \to \mathcal H$.
As we have seen earlier, to formulate the associated effective theory in the Schwinger--Keldysh framework, the symmetry group 
is extended as
\begin{equation}
\widetilde{\mathcal G} \coloneqq \mathcal G_1 \times \mathcal G_2 ,
\end{equation}
where $\mathcal G_1$ and $\mathcal G_2$ act independently on the forward and
backward time branches of the Schwinger--Keldysh contour, respectively.
The extended symmetry acts on the density matrix as
\begin{equation}
\rho \;\mapsto\; U(g_1)\,\rho\,U^\dagger(g_2),
\end{equation}
with $g_1 \in \mathcal G_1$ and $g_2 \in \mathcal G_2$.

The connection to strong and weak symmetries is then immediate.
A strong symmetry corresponds to an independent action of $\mathcal G_1$
or $\mathcal G_2$ on the density matrix, while a weak symmetry is associated
with
\changed{%
the diagonal subgroup
\begin{equation}
\mathcal G_r
\;\subset\;
\mathcal G_1 \times \mathcal G_2,
\end{equation}
}%
which acts identically on both time branches.
We can summarize the correspondence as
\changed{%
\begin{subequations}\begin{align}
\text{strong symmetry  $G$} \quad &\longleftrightarrow \quad \mathcal G_1 \text{ or } \mathcal G_2, \\
\text{weak symmetry   $G$} \quad &\longleftrightarrow \quad \mathcal G_r.
\end{align}\end{subequations}
}%

When a thermal state is realized, the density matrix does not remain invariant
under independent transformations of $\mathcal G_1$ and $\mathcal G_2$.
Instead, only the diagonal subgroup leaves the thermal state invariant.
As a result, the symmetry breaking pattern in the Schwinger--Keldysh framework
takes
\changed{%
the form
\begin{equation}
\widetilde{\mathcal G}
\;\longrightarrow\;
\mathcal H_r,
\end{equation}
where $\mathcal H_r$ denotes
}%
the diagonal subgroup of $\mathcal H_1 \times \mathcal H_2$
 (i.e., $\rho \mapsto U(h) \rho U^\dag(h)$ for $h \in \mathcal H$).

Even when the symmetry $\mathcal G$ itself is unbroken,
the Schwinger--Keldysh formulation naturally leads to
\changed{%
a nontrivial coset
structure,
\begin{equation}
\widetilde{\mathcal G} \;\longrightarrow\; \mathcal G_r,
\end{equation}
}%
reflecting the fact that a thermal state preserves only the diagonal subgroup
of the extended symmetry.

This reduction of symmetry can be understood as an instance of
strong-to-weak symmetry breaking~\cite{Lessa:2024gcw, Gu:2024wgc, Huang:2024rml}, in which a strong symmetry of the
time evolution is realized only as a weak symmetry at the level of the
effective dynamics.
Furthermore, insights from the Schwinger--Keldysh coset construction
provide a systematic recipe for constructing effective field theories
associated with a generic symmetry breaking pattern
$\mathcal G \to \mathcal H$.

As discussed in section~\ref{sec:origin-diffusive-symmetry}, the fact that the
relevant coset manifold is
\changed{%
given by $\widetilde{\mathcal G}/\mathcal H_r$
}%
implies that, even in the unbroken phase, the effective theory must exhibit a
residual local invariance associated with the diagonal generators.
This residual invariance constitutes the origin of the diffusive symmetry.

\section{Summary and discussions}\label{sec:summary}

In this work, we developed a symmetry-based effective field theory for photons in insulating media at finite temperature by combining the generalized coset construction for higher-form symmetries with the Schwinger--Keldysh formalism. Within this framework, the photon is
identified as the Nambu--Goldstone mode associated with the spontaneous breaking of the
$1$-form symmetry ${\rm U}(1)^{[1]}_+ \times {\rm U}(1)^{[1]}_- \to 1$, and the doubled gauge
fields naturally arise as the effective degrees of freedom.  
We constructed the most general low-energy effective action up to $m+n \le 4$ consistent with gauge invariance, rotational symmetry, unitarity, and the dynamical KMS symmetry. At this order, dissipation appears through two transport coefficients, $\tau_E$ and $\tau_B$.

We demonstrated that the Schwinger--Keldysh action is equivalent to a set of Langevin-type stochastic Maxwell equations, with the noise terms obeying the fluctuation--dissipation relation. We also analyzed several physical consequences of the effective theory. We identified the entropy current from the Schwinger--Keldysh action, which satisfies a non-negative divergence.
Furthermore, we examined electromagnetic duality in this dissipative setting.

Our results provide a model-independent and symmetry-based description of thermal photon dynamics in insulating media. Since the effective theory is organized as a systematic derivative expansion, the description can be systematically improved by incorporating higher-order terms. The construction demonstrates that higher-form symmetries serve as a powerful organizing principle not only at zero temperature but also at finite temperature, where dissipative
effects become important.

We also discussed the effective field theory for an unbroken ${\rm U}(1)$ $1$-form symmetry. In this phase, the low-energy dynamics is governed by diffusive modes rather than propagating excitations, and we clarified that the associated diffusive shift symmetry originates from a residual gauge redundancy that remains after fixing the thermodynamic parameters in the Schwinger--Keldysh formulation. We further related the Schwinger--Keldysh symmetry structure to the notions of strong and weak symmetries discussed in the context of open quantum systems.

Let us conclude by commenting on several possible future directions. A natural extension of the present work is to consider media with additional broken spacetime symmetries. Examples include systems lacking spatial rotational or translational symmetry, where the reduced symmetry allows for new transport coefficients and gapless degrees of freedom. These changes lead to qualitatively new dissipative responses of gauge fields.

Examples include systems lacking spatial rotational or parity symmetry, where the reduced symmetry allows for new transport coefficients and leads to qualitatively new dissipative responses of gauge fields.

Another important direction is to investigate the effects of explicit symmetry breaking in higher-form dissipative effective field theories. For ordinary $0$-form symmetries, the interplay between explicit breaking, dissipation, and collective modes has been studied extensively~\cite{Hidaka:2012ym,Hongo:2024brb}. Extending such analyses to higher-form symmetries would clarify how finite symmetry-breaking scales modify diffusive photon dynamics and relaxation
processes in insulating media.

It would also be interesting to explore how the interplay between explicit symmetry breaking and dissipation modifies the dynamics of Nambu--Goldstone modes. For spontaneously broken internal 0-form symmetries, Nambu--Goldstone modes are classified into type-A and type-B, depending on whether the broken generators form canonical pairs~\cite{Nielsen:1975hm,Watanabe:2012hr,Hidaka:2012ym}. Type-A modes exhibit linear dispersion relations, while type-B modes display quadratic dispersions as a consequence of their nontrivial commutation relations. In the presence of higher-form symmetries, the interplay between explicit breaking, dissipation, and type-B structures is expected to further modify these dispersion relations, giving rise to quadratic dispersion relations and the coexistence of propagating and diffusive collective modes~\cite{Sogabe:2019gif,Hidaka:2020ucc}.
It would be interesting to study these phenomena from the viewpoint of the Schwinger--Keldysh effective theory.

Finally, it would be interesting to extend the present framework to gauge theories hosting fractonic excitations. Fractons~\cite{annurev:/content/journals/10.1146/annurev-conmatphys-040721-023549,Gromov:2022cxa} are collective excitations with restricted mobility and have been studied extensively in condensed matter physics. It is well established that higher-rank gauge theories naturally realize such restricted worldline dynamics~\cite{Pretko:2017kvd,Pretko:2016kxt,Pretko:2016lgv,Perez:2023uwt}.
More recently, it has been shown that these theories can be understood as arising from the spontaneous breaking of higher-form symmetries whose conserved charges do not commute with spatial translations, with the resulting symmetry algebra directly dictating the mobility constraints of excitations~\cite{Hirono:2022dci}. It would therefore be interesting to extend the Schwinger--Keldysh effective field theory framework to such nonuniform higher-form symmetries, enabling a systematic treatment of dissipative effects in fractonic gauge theories.

\acknowledgments
We are grateful to Noriyuki Sogabe for useful discussions.
Y.\,A. also thanks Masaru Hongo for fruitful communications.
G.\,Y. is supported by the Kondo Memorial Foundation and the Honors Program for Graduate Schools in Science, Engineering and Informatics, the University of Osaka.
Y.\,A. is supported by Japan Society for the Promotion of Science (JSPS) KAKENHI Grant Numbers JP23H01174 and JP23K25870.
Y.\,H. is supported in part by JSPS KAKENHI Grant Numbers JP22H05111, JP22H05118, JP24K23186, and by JST, PRESTO Grant Number JPMJPR24K8.

\appendix

\section{Terms invariant up to total derivatives under 1-form symmetry}\label{app:ChernSimonsForm}

In this appendix, we classify terms that are invariant under $\mathrm{U}(1)^{[1]}$ transformations up to total derivatives. We show that, in $D$-dimensions, the only such terms that cannot be written solely in terms of the field strengths $f$ and $F$ take the form\footnote{
    We use the notation $(F)^n \coloneqq \underbrace{F\wedge\cdots\wedge F}_{n}$.
}
\begin{equation}
    \Ocal[a,A] =
    \left\{\begin{aligned}
        a \wedge (f)^n \wedge (F)^m
        \quad
        &\text{where}\;\;D=2(n+m)+1,\\
        u\wedge a \wedge (f)^n \wedge (F)^m
        \quad
        &\text{where}\;\;D=2(n+m)+2,
    \end{aligned}\right.
    \label{CSformSK}
\end{equation}
which are invariant under $\mathrm{U}(1)^{[1]}_+\times\mathrm{U}(1)^{[1]}_-$ transformations up to total derivatives, but cannot be expressed purely in terms of $f$ and $F$. 
We first establish this result for a single $\mathrm{U}(1)^{[1]}$ symmetry,
and then extend it to the $\mathrm{U}(1)^{[1]}_+\times\mathrm{U}(1)^{[1]}_-$
symmetry required in the Schwinger--Keldysh effective action.

We first show that local terms containing two factors of $A$ cannot be invariant
under the $\mathrm{U}(1)^{[1]}$ symmetry.
If such a term existed, it would generically take the form
\begin{equation}
    \Ocal[A]
    = \sum_{n=0}^N
        A_\mu
        (\p_{\rho_1}\cdots\p_{\rho_n}A_\nu)
        M_n^{\mu\nu\rho_1\cdots\rho_n}[F].
    \label{ExpectedA^2term}
\end{equation}
The tensor $M^{\mu\nu\rho_1\cdots\rho_n}_n$ must be totally symmetric;
other wise, the assumption that the term contains two explicit factors of $A$
would be violated.
We also require that $\sum M_n[F](\partial)^n$ is Hermitian, i.e., after performing partial integration $n$ times for each term in $\Ocal[A]$, it is still the same up to total derivatives.
We can always make it Hermitian by subtracting the anti-Hermitian part, which only contributes to total derivatives.

The Hermiticity condition implies that $N$ must be even and $M_N^{\mu\nu\rho_1\cdots\rho_n}$ is a totally symmetric tensor (including the first index $\mu$) as shown below.
For terms containing $N$ derivatives on $A$, it leads to
\begin{align}
    M_N^{\mu\nu\rho_1\cdots\rho_N}
    =(-1)^N M_N^{\nu\mu\rho_1\cdots\rho_N}.
    \label{Hermitian_N}
\end{align}
When $N$ is odd, $M_N^{\mu\nu\rho_1\cdots\rho_N}$ must be anti-symmetric under $\mu\leftrightarrow \nu$ and symmetric under $\nu\leftrightarrow\rho_1$, which is impossible because
\begin{align}
    M_N^{\mu\nu\rho_1\cdots}
    = M_N^{\mu\rho_1\nu\cdots}
    = -M_N^{\rho_1\mu\nu\cdots}
    = (-1)^2 M_N^{\nu\rho_1\mu\cdots}
    = (-1)^3 M_N^{\mu\nu\rho_1\cdots} = 0.
    \label{AntisymVanish}
\end{align}
Therefore, $N$ must be even, and in this case $M_N^{\mu\nu\rho_1\cdots\rho_N}$ is a totally symmetric tensor. 

In order for $\Ocal[A]$ to be $\mathrm{U}(1)^{[1]}$-symmetric, its variation under the infinitesimal transformation, $A \to A + \Lambda$ satisfying $\rd\Lambda=0$, must vanish up to total derivatives.
Using the Hermiticity property, it requires
\begin{equation}
 0=\Lambda_\mu\sum_{n=0}^{N}(\p_{\rho_1}\cdots\p_{\rho_n}A_\nu)
        M_n^{\mu\nu\rho_1\cdots\rho_n}[F],
\end{equation}
for arbitrary $\Lambda$ and $A$.
It implies the existence of a conserved current for arbitrary $A$:
\begin{equation}
 0=\partial_\mu\sum_{n=0}^{N}(\p_{\rho_1}\cdots\p_{\rho_n}A_\nu)
        M_n^{\mu\nu\rho_1\cdots\rho_n}[F].
\end{equation}
Redundancy of $A$ for a fixed $F$ enables us to derive separate conditions for the coefficients of $\partial_{\mu}\partial_{\rho_1}\cdots\partial_{\rho_n}A_{\nu}$.
For $n=N$, it leads to $M_N^{\mu\nu\rho_1\cdots\rho_N}=0$ because there is no anti-symmetric pairs in the indices of $M_N$, which would satisfy the condition automatically.

If we assume $\Ocal[A]$ to be local, there must exist a maximum order of derivatives on $A$ for each $\Ocal[A]$.
However, the above argument has derived $M_N^{\mu\nu\rho_1\cdots\rho_N}=0$, in contradiction to the fact that $N$ is the maximum order.
Therefore, a local term $\Ocal[A]$ which contains two $A$'s and is invariant under $\mathrm U(1)^{[1]}$ does not exist.

Next, we consider a term containing only one $A$. Terms linear in $A$ can be expressed using a $1$-form $C_1[F]$:
\begin{equation}
    \Ocal[A] = A \wedge \star C_1[F].
\end{equation}
The $\mathrm{U}(1)^{[1]}$ symmetry requires that the change of $\Ocal[A]$ under the transformation, $A\to A + \Lambda$ with $\rd\Lambda=0$, should vanish up to total derivatives.
It implies
\begin{equation}
    \Lambda \wedge \star C_1[F] = \Lambda\wedge\rd\star K_2[A] = -\rd (\Lambda\wedge\star K_2[A]),
\end{equation}
for arbitrary closed $1$-form $\Lambda$, and thus $\star C_1[F]$ is exact
\begin{align}
    \star C_1[F] = \rd\star K_2[A].
    \label{exact_c1}
\end{align}
If the 2-form $K_2$ depends only on $F$, i.e., $K_2=K_2[F]$, we would get
\begin{equation}
    \Ocal[A] = A \wedge \rd\star K_2[F] = F \wedge \star K_2[F] - \rd(A \wedge \star K_2[F]),
\end{equation}
leading to $\Ocal=\Ocal[F]$ up to total derivatives, which contradicts our assumption.
Thus, $K_2$ must depend explicitly on $A$ but $\rd\star K_2[A]$ does not.

Let us construct $K_2[A]$ which depends explicitly on $A$ while $\rd\star K_2[A]$ depends only on $F=\rd A$.
Such $\star K_2[A]$ must contain only one $A$.
If it contains more than one $A$'s, those terms would vanish under the exterior derivative $\rd$ and thus is unnecessary for the construction of $\Ocal[A]$.
These considerations lead to the most general form of $\star K_2[A]$ as
\begin{equation}
    \star K_2[A] = A \wedge \star C_3[F], \quad \rd\star C_3[F] = 0,
    \label{k2=ac3}
\end{equation}
using a co-closed 3-form $C_3[F]$.
We will prove this statement later using tensor analysis.
Now we have
\begin{align}
    \Ocal[A] 
    = A \wedge \star C_1[F]
    = A \wedge \rd (A \wedge \star C_3[F])
    = A \wedge F \wedge \star C_3[F].
\end{align}
By the Poincar\'e's lemma, $\star C_3[F]$ can be a constant or exact $\star C_3[F] = \rd\star K_{4}[A]$.
If the latter is the case, we can repeat a recursive manipulation starting from eq.~\eqref{exact_c1} to arrive at a co-closed 5-form $C_5[F]$.
This loop continues until a closed $D-(2n+1)$-form $\star C_{2n+1}[F]$ is a constant.
For example, a constant 0-form is 1 for $D=2n+1$ and a constant $1$-form can be $u=u_{\mu} \rd x^{\mu}=-\rd t$ in the medium rest frame for $D=2n+2$.
In summary, the resulting non-trivial terms $\Ocal[A]$ take the following forms:
\begin{equation}
    \Ocal[A] =
    \left\{\begin{aligned}
        A\wedge (F)^n
        \quad
        &\text{when}\;\;D=2n+1\\
        u\wedge A\wedge (F)^n
        \quad
        &\text{when}\;\;D=2n+2.
    \end{aligned}\right.
    \label{CSformMicro}
\end{equation}

We will now prove that if $2n$-form $K_{2n}[A]$ contains only one $A$ while $\rd\star K_{2n}[A]$ depends only on $F=\rd A$, it must be written as ($n=1$ corresponds to eq.~\eqref{k2=ac3})
\begin{align}
    \star K_{2n}[A] = A \wedge \star C_{2n+1}[F], \quad
    \rd \star C_{2n+1}[F] = 0 .
    \label{keven_acodd}
\end{align}

\begin{proof}
General form of $\star K_{2n}[A]$ containing only one $A$ is
\begin{equation}
    K_{2n}^{\mu_1\cdots\mu_{2n}}[A]
    = \sum_{m=0}^M
        (\p_{\rho_1}\cdots\p_{\rho_m}A_{\nu})
        M_{m}^{\mu_1\cdots\mu_{2n},\nu\rho_1\cdots\rho_m}[F],
\end{equation}
where the indices $\mu_1,\cdots,\mu_{2n}$ are anti-symmetric in both $K_{2n}^{\mu_1\cdots\mu_{2n}}$ and $M_m^{\mu_1\cdots\mu_{2n},\nu\rho_1\cdots\rho_m}$ while the indices $\nu,\rho_1,\cdots,\rho_m$ are symmetric in $M_m^{\mu_1\cdots\mu_{2n},\nu\rho_1\cdots\rho_m}$ because there is an explicit $A$ by assumption.
Since $\rd\star K_{2n}[A]$ depends only on $F$, it follows that
\begin{align}
    \partial_{\mu_1}K_{2n}^{\mu_1\cdots\mu_{2n}}[A+\Lambda]
    -\partial_{\mu_1}K_{2n}^{\mu_1\cdots\mu_{2n}}[A]
    = \partial_{\mu_1}\sum_{m=0}^M
     (\p_{\rho_1}\cdots\p_{\rho_m}\Lambda_{\nu})
     M_{m}^{\mu_1\cdots\mu_{2n},\nu\rho_1\cdots\rho_m}[F]
    =0.
\end{align}
It holds for arbitrary closed $1$-form $\Lambda$ and separate conditions can be derived for coefficients of $\p_{\mu_1\rho_1\cdots\rho_m}\Lambda_{\nu}$.
For $m=M$, it implies that $M_{M}^{\mu_1\cdots\mu_{2n},\nu\rho_1\cdots\rho_M}$ must be anti-symmetric under $\mu_1\leftrightarrow\nu$ because $\p_{\mu}\Lambda_{\nu} - \p_{\nu}\Lambda_{\mu}=0$.
When $M\geq 1$, it is symmetric under $\nu\leftrightarrow\rho_1$ so that we can show $M_{M}^{\mu_1\cdots\mu_{2n},\nu\rho_1\cdots\rho_M}=0$ by a similar mechanism with eq.~\eqref{AntisymVanish}.
Then, $M=0$ and $M_0^{\mu_1\cdots\mu_{2n},\nu}$ is a totally anti-symmetric tensor and is conserved $\p_{\nu}M_0^{\mu_1\cdots\mu_{2n},\nu}[F]=0$.
This implies that a co-closed $(2n+1)$-form $C_{2n+1}[F]$ can be constructed by $M_0^{\mu_1\cdots\mu_{2n},\nu}$ such that:
\begin{align}
    \star K_{2n}[A] = A \wedge \star C_{2n+1}[F], \quad
    \rd \star C_{2n+1}[F] = 0.
\end{align}
\end{proof}

Finally, we explain how this result extends to the Schwinger--Keldysh formalism, where the symmetry is doubled as $\mathrm{U}(1)^{[1]}_r\times\mathrm{U}(1)^{[1]}_a$. 
If $\mathrm{U}(1)^{[1]}_r$ and $\mathrm{U}(1)^{[1]}_a$ symmetries
are considered independently, the discussion above implies that $\Ocal[a,A]$ must be at most linear in $A$ and $a$, which allows a bilinear term $a\wedge A\wedge C[f,F]$.
However, $\mathrm U(1)^{[1]}_+$ symmetry disallows such terms, 
and $\Ocal[a,A]$ must be at most linear in $A$ {\it or} $a$.
In the derivation above, we just have to replace 
eq.~\eqref{keven_acodd} with the following:
\begin{align}
    \star K_{2n}[a,A]
    = (a \ \text{or} \ A) \wedge \star C_{2n+1}[f,F], \quad
    \rd \star C_{2n+1}[f,F] = 0 .
\end{align}
As a result, a part of $A,F$ in eq.~\eqref{CSformMicro} is replaced by $a,f$. Because of the unitarity condition~\eqref{Unitarity1}, $\Ocal[a,A]$ is at least of order $a^1$, so it can always be written in the form eq.~\eqref{CSformSK}, up to total derivatives.

\section{Chiral magnetic effect in the Schwinger--Keldysh EFT}\label{sec:CME}

Here, we examine the term $\bm a \cdot \bm B$ and clarify its relation to the Chiral Magnetic Effect (CME)~\cite{Kharzeev:2007jp, Fukushima:2008xe}.
The CME refers to the generation of an electric conduction current
along a magnetic field in the presence of a chirality imbalance,
\begin{equation}
  \bm J_\mathrm{cond} \propto \mu_5\,\bm B ,
\end{equation}
where $\mu_5$ denotes the axial chemical potential.
It is a macroscopic transport phenomenon whose microscopic origin lies in the chiral anomaly.
The CME can also be described within the present framework of non-equilibrium effective field theory. Here, we demonstrate that the CME is encoded in a Chern--Simons--like term in the effective action, and discuss the conditions under which it is allowed.

In systems with dynamical charged matter, the CME manifests itself as an electric conduction current. In such cases, the presence of charged matter explicitly breaks the $\mathrm{U}(1)_+^{[1]}\times \mathrm{U}(1)_-^{[1]}$ $1$-form symmetry, and the conduction current enters the modified conservation law for the 2-form current,
\begin{equation}
    \rd\star J^{(2)} = \star J_\mathrm{cond}^{(1)}.
\end{equation}
In contrast, in insulating systems where mobile charge carriers are absent,
it is more appropriate to focus on an induced current. 
This current originates from the collective motion of electric dipoles
and is commonly discussed in the context of polarization or magnetization responses. In our effective description, the induced current is defined by 
\begin{equation}
     \rd\star J^{(2)} = 0, \quad
     \star J_\mathrm{ind}^{(1)}
     \coloneqq
     \rd\star (J^{(2)}-F^{(2)}).
\end{equation}
Although our EFT does not explicitly include additional dipole degrees of freedom, such effects can be effectively incorporated through their coupling to electromagnetic fields, leading to dressed photon degrees of freedom.
Within this framework, the chiral magnetic response is encoded in the induced current.
Now, consider the following term in the effective Lagrangian,
\begin{equation}
\bm a \cdot \bm B.   
\label{eq:a-dot-B}
\end{equation}
This term gives rise to an induced current proportional to the magnetic field,
\begin{equation}
    {\bm J}_\mathrm{ind} \propto 
    \bm{B},
\end{equation}
which can be deduced by noting that the conservation law \eqref{ConservedCurr} of 2-form current $J$ comes from the variation $\delta I_{\rm eff}/\delta a$.
Thus, we can see the term~\eqref{eq:a-dot-B} encodes the CME current in the Schwinger-Keldysh EFT.

We note, however, that the term \eqref{eq:a-dot-B} does not lead to a
gauge-invariant current operator, as we explain below.
If this term is included, the effective Lagrangian can be written as
\begin{equation}
\Lcal'[\rd a, \rd A] + c\, a \wedge u \wedge F,
\end{equation}
where $\Lcal'[\rd a, \rd A]$ denotes a Lagrangian that is strictly
invariant under the $1$-form symmetries (without generating total-derivative
terms), 
$c$ is a constant, 
and $u = u_\mu \rd x^\mu$ is a constant $1$-form with $u^\mu = (1, \bm 0)$.
The second term reduces to $\bm a \cdot \bm B$ in components.

Varying this action with respect to $a$ yields the Ward--Takahashi identity
\begin{equation}
\rd \star J' = - c\, u \wedge F ,
\end{equation}
where $J'$ is the $2$-form current derived from $\Lcal'$.
Formally, this relation can be rewritten as a conservation law,
\begin{equation}
\rd \left( \star J' - c\, u \wedge A \right) = 0 .
\end{equation}
However, the corresponding conserved current necessarily contains the gauge
potential $A$ without derivatives and is therefore gauge-variant.
This demonstrates that the term~\eqref{eq:a-dot-B} does not lead to a
gauge-invariant current associated with the $1$-form symmetry.

From here, we examine under what conditions the term~\eqref{eq:a-dot-B}
is compatible with the dynamical KMS symmetry.
We first note that this term must be non-dissipative. If it were dissipative, it would require a dynamical KMS partner term, which is higher order in the $a$ field and lower order in time derivatives. However, since the term~\eqref{eq:a-dot-B} does not contain any time derivative, no such dynamical KMS partner exists.

Therefore, if the term~\eqref{eq:a-dot-B} is present in the effective Lagrangian, it must be $\Theta$-even, provided that the $\Theta$ symmetry is not spontaneously broken. 
When $\Theta = \mathcal{T},\ \mathcal{CT}$, this term is $\Theta$-even.
One can explicitly verify that it is invariant under the dynamical KMS transformation up to total derivatives:
\begin{equation}
    \mathrm{T}_\Theta\cdot\mathrm{KMS}_\Theta
        [\bm{a}\cdot\bm{B}]
        - \bm{a}\cdot\bm{B}
    = \im\beta\dot{\bm A}\cdot\bm{B}
    = -\im\beta \bm{E}\cdot\bm{B}
        - \im\beta \bm\nabla\cdot(A_0\bm{B}),
\end{equation}
where $\bm{E}\cdot\bm{B}$ a total derivative.
When $\Theta=\mathcal{PT,\ CPT}$, the term~\eqref{eq:a-dot-B} is $\Theta$-odd.
However, it can still be made consistent with the dynamical KMS symmetry by multiplying it with a $\Theta$-odd coefficient. Such a coefficient is allowed only when the $\Theta$ symmetry is spontaneously broken.

Finally, we discuss the instability induced by the term $\bm a \cdot \bm B$. In the presence of this term, the leading contributions to the effective Lagrangian are given by
\begin{equation}
    \epsilon\,\bm e \cdot \bm E
    + \chi\,\bm a \cdot \bm B .
\end{equation}
Here we have retained the terms that are lowest order in the $a$ field
and involve the smallest number of temporal or spatial derivatives.
Using the Bianchi identity \eqref{Bianchi}, the electric field $\bm E$ can be eliminated from the equations of motion,
leading to
\begin{equation}
    - \epsilon\,\ddot{\bm B}
    + \chi\,\bm\nabla \times \bm B
    = \bm 0 .
\end{equation}
This equation admits two modes with opposite helicities,
\begin{equation}
    \omega(k)
    = \pm \im \sqrt{\frac{\chi}{\epsilon}} \sqrt{k} .
\end{equation}
One of these modes grows exponentially in time, signaling an instability.
The instability of dynamical electromagnetic fields in the presence of a finite chirality imbalance,
as well as the subsequent formation of an inverse cascade,
has been extensively studied in literature~\cite{Joyce:1997uy,Akamatsu:2013pjd,Boyarsky:2011uy,Tashiro:2012mf,Hirono:2015rla,Yamamoto:2016xtu,Hattori:2017usa}.

In summary, the term $\bm a \cdot \bm B$ in the effective Lagrangian encodes the
chiral magnetic effect within the Schwinger--Keldysh EFT framework.
This term is compatible with the dynamical KMS condition when either
$\mathcal{PT}$ or $\mathcal{CPT}$ symmetry is explicitly or spontaneously
broken.
However, the associated $1$-form currents are necessarily gauge-variant, and
the presence of this term generically leads to an instability of the
electromagnetic fields.
Understanding the eventual fate of this instability requires going beyond
the linearized analysis and incorporating nonlinear effects.

\section{Details on dynamical KMS symmetry}\label{app:dkms}

This appendix collects technical details on the dynamical KMS transformation used in the main text. 

\subsection{Dynamical KMS transformation of fields}\label{app:DKMS-Calc}

Let us here discuss the transformation properties of Abelian gauge fields, which are building blocks of the effective Lagrangian for spontaneously broken higher-form symmetries.
The key input is the action of an anti-unitary
$\mathbb{Z}_2$ symmetry 
$\Theta$ of the underlying microscopic theory on the doubled gauge fields $\{A,a\}$. Because we work in a non-relativistic setting, the temporal and spatial components of the $(p+1)$-form field strength (equivalently, the Maurer--Cartan form) must be treated separately. 

We begin by determining the most general $\Theta$-transformation of the gauge
potential consistent with $F=\rd A$ and gauge redundancy (allowing for total
derivatives). The resulting transformation rule will then be used as the input for the dynamical KMS transformation discussed later in this appendix.

The basic building block of the effective Lagrangian is the field strength
$F^{(p+1)}$. Under an anti-unitary symmetry, observables may pick up an overall sign. Accordingly, we parameterize the $\Theta$-parities of the temporal and spatial components of $F$ by
\begin{equation}
    \left\{\begin{aligned}
        \mathrm{T}_\Theta\cdot\Theta
            [F_{0 i_1\cdots i_p}]
        &= \eta_E F_{0 i_1\cdots i_p},\\
        \mathrm{T}_\Theta\cdot\Theta
            [F_{ i_1\cdots i_{p+1}}]
        &= \eta_B F_{ i_1\cdots i_{p+1}},
    \end{aligned}\right.
\end{equation}
with $\eta_{E,B} = \pm 1$.
Since $F=\rd A$, a compatible transformation of the gauge field can be written as\footnote{
    The condition $\mathrm{T}_\Theta\cdot\Theta[F]-\eta_{E,B}F=0$ can be viewed as a closure condition in a $(p+1)$-dimensional hyperplane.
    In the case $p=1$, $\alpha$ and $\gamma$ are
    0-forms; for instance, $\p_{[0}\alpha_{ i_1\cdots i_{p-1}]}=\dot\alpha$.
}
\begin{equation}
    \left\{\begin{aligned}
        \mathrm{T}_\Theta\cdot\Theta
            [A_{0 i_1\cdots i_{p-1}}]
        &= \eta_E\eta (A_{0 i_1\cdots i_{p-1}}
            + \p_{[0}\alpha_{ i_1\cdots i_{p-1}]}(t,\bm x)),\\
        \mathrm{T}_\Theta\cdot\Theta
            [A_{ i_1\cdots i_p}]
        &= -\eta_E (A_{ i_1\cdots i_p}
            - \p_{[ i_p}\alpha_{ i_1\cdots i_{p-1}]}(t,\bm x)
            - \beta_{ i_1\cdots i_p}(\bm x))\\
        &= \eta_B\eta (A_{ i_1\cdots i_p}
            + \p_{[ i_p}\gamma_{ i_1\cdots i_{p-1}]}(t,\bm x)),
    \end{aligned}\right.
\end{equation}
where $\alpha$ and $\gamma$ are $(p-1)$-forms and $\beta$ is a $p$-form, which may depend on the fields $\{a,A\}$.

If one takes $\eta_E\eta_B=\eta$, the above relations require
\begin{equation}
    2A_{ i_1\cdots i_p}(t,\bm x)
    = \p_{[ i_p}\alpha_{ i_1\cdots i_{p-1}]}(t,\bm x)
        + \beta_{ i_1\cdots i_p}(\bm x)
        - \p_{[ i_p}\gamma_{ i_1\cdots i_{p-1}]}(t,\bm x).
\end{equation}
However, this condition cannot be satisfied in general.
We therefore conclude that consistency of the $\Theta$ symmetry with gauge
redundancy requires
\begin{equation}
    \eta_E\eta_B=-\eta
\end{equation}
In this case the choice of $\alpha$ (equivalently $\gamma$) can be absorbed into a gauge transformation, and the $\Theta$ action on $A$ simplifies to
\begin{equation}
    \left\{\begin{aligned}
        \Theta A_{0 i_1\cdots i_{p-1}}
        &= \eta_A A_{0 i_1\cdots i_{p-1}},\\
        \Theta A_{ i_1\cdots i_p}
        &= -\eta_A\eta A_{ i_1\cdots i_p},
    \end{aligned}\right.
\end{equation}
with 
\begin{equation}
    \eta_A \coloneqq \eta_E\eta = -\eta_B.
\end{equation}
For $p=1$ (i.e., electromagnetism), $\eta_A=+1$ for $\Theta = \mathcal{T, PT}$ and $\eta_A=-1$ for $\Theta = \mathcal{CT, CPT}$. 
The fluctuation field $a$ should transform in the same way:
\begin{equation}
    \left\{\begin{aligned}
        \Theta a_{0 i_1\cdots i_{p-1}}
        &= \eta_A a_{0 i_1\cdots i_{p-1}},\\
        \Theta a_{ i_1\cdots i_p}
        &= -\eta_A\eta a_{ i_1\cdots i_p}.
    \end{aligned}\right.
\end{equation}

We now turn to the dynamical KMS transformation.
Once the action of the anti-unitary symmetry $\Theta$ has been fixed, the
corresponding dynamical KMS transformation follows straightforwardly.
It is given by
\begin{equation}
    \left\{\begin{aligned}
        \tilde A_{0 i_1\cdots i_{p-1}}(t,\bm x)
        &= \eta_A A_{0 i_1\cdots i_{p-1}}(-t,\eta \bm x),\\
        \tilde A_{ i_1\cdots i_p}(t,\bm x)
        &= -\eta_A\eta A_{ i_1\cdots i_p}(-t,\eta \bm x),\\
        \tilde a_{0 i_1\cdots i_{p-1}}(t,\bm x)
        &= \eta_A [a_{0 i_1\cdots i_{p-1}}(-t,\eta \bm x)
            + \im\beta \p_{-t} A_{0 i_1\cdots i_{p-1}}(-t,\eta \bm x)],\\
        \tilde A_{ i_1\cdots i_p}(t,\bm x)
        &= -\eta_A\eta [a_{ i_1\cdots i_p}(-t,\eta \bm x)
            + \im\beta \p_t A_{ i_1\cdots i_p}(-t,\eta \bm x)],
    \end{aligned}\right.
\end{equation}
where $\p_{-t} \coloneqq \p/\p(-t)$ and later we also use $\bm\nabla_{\eta x} \coloneqq \p/\p(\eta\bm{x})$. 
The $\Theta$ transformation acts on spacetime coordinates as
\begin{equation}
    \Theta\p_t
    = \p_{-t}
    = -\p_t,
    \quad
    \Theta\bm\nabla_x
    = \bm\nabla_{\eta x}
    = \eta\bm\nabla_x,
\end{equation}
whereas the dynamical KMS transformation acts only on the field operators.
Since $\Theta$ is a $\mathbb{Z}_2$ symmetry,
the dynamical KMS transformation is also of
$\mathbb{Z}_2$ type.

Let us demonstrate the transformation explicitly 
for the case $\Theta = \mathcal{PT}$ or $\mathcal{CPT}$. 
In this case, the temporal and spatial components of $a, A$ obey the same transformation rule under $\Theta$, and we have
\begin{equation}
    \begin{aligned}
        &({\mathrm{KMS}_{\Theta}})^2[A(t,\bm x)]\\
        &= \eta_A \tilde{A}(-t,-\bm x)\\
        &= \eta_A^2 A(t,\bm x)\\
        &= A(t,\bm x),
    \end{aligned}
    \qquad\qquad
    \begin{aligned}
        &({\mathrm{KMS}_{\Theta}})^2[a(t,\bm x)]\\
        &= \eta_A \tilde{a}(-t,-\bm x)
            + \im\eta_A \beta\p_{-t}\tilde{A}(-t,-\bm x)\\
        &= \eta_A^2 a(t,\bm x)
            + \im\beta\eta_A^2 \p_{t}A (t,\bm x)
            + \im\eta_A^2 \beta\p_{-t} A(t,\bm x)\\
        &= a(t,\bm x).
    \end{aligned}
\end{equation}
The $\mathbb{Z}_2$ nature
can be verified in a similar manner for $\Theta=\mathcal{T}$ or $\Theta=\mathcal{CT}$. 
In these cases, however, one must distinguish whether the indices of the
$p$-form gauge field $A^{(p)}$ include a temporal component, and the analysis
must be carried out separately for each case.

For $p=1$ and $D=3+1$, the electric and magnetic fields are transformed under general $\Theta$ as
\begin{subequations}
\begin{align}
    &\begin{aligned}
        \mathrm{T}_\Theta\cdot\mathrm{KMS}_\Theta
            [\bm{E}(t,\bm x)]
        &=  \mathrm{T}_\Theta[
                - \bm{\nabla}_x\tilde{A}_0 (t,\bm x) 
                - \p_t\tilde{\bm{A}} (t,\bm x)
            ]\\
        &=  \mathrm{T}_\Theta[
                - \eta_A\bm{\nabla}_x A_0 (-t,\eta \bm x) 
                + \eta_A\eta\p_t \bm{A} (-t,\eta \bm x)
            ]\\
        &=  - \eta_A\bm{\nabla}_{\eta x} A_0 (t,\bm x) 
            + \eta_A\eta\p_{-t} \bm{A} (t,\bm x)\\
    &=  \eta_A\eta \bm{E}(t,\bm x),
    \end{aligned}\\
    &\begin{aligned}
        \mathrm{T}_\Theta\cdot\mathrm{KMS}_\Theta
            [\bm{B} (t,\bm x)]
        &=  \mathrm{T}_\Theta\cdot\mathrm{KMS}_\Theta
                [\bm\nabla_x\times\bm{A} (t,\bm x)]\\
        &=  \mathrm{T}_\Theta
                [-\bm\nabla_x\times\eta_A\eta\bm{A} (-t,\eta \bm x)]\\
        &=  - \bm\nabla_{\eta x}\times\eta_A\eta\bm{A} (t,\bm x)\\
        &=  - \eta_A \bm{B}(t,\bm x),
    \end{aligned}
\end{align}
\begin{align}
    &\begin{aligned}
        \mathrm{T}_\Theta\cdot\mathrm{KMS}_\Theta
            [\tilde{\bm{e}}(t,\bm x)]
        &=  \mathrm{T}_\Theta\cdot\mathrm{KMS}_\Theta[
                - \bm{\nabla}_x\tilde{a}_0 (t,\bm x) 
                - \p_t\tilde{\bm{a}} (t,\bm x)
            ]\\
        &=  \mathrm{T}_\Theta[
                - \eta_A\bm{\nabla}_x a_0 (-t,\eta \bm x) 
                + \eta_A\eta \p_t \bm{a} (-t,\eta \bm x)\\
        &\qquad\qquad
                - \im\eta_A\beta\bm{\nabla}_x \p_{-t} A_0 (-t,\eta \bm x) 
                + \im\eta_A\eta\beta\p_t \p_{-t} \bm{A} (-t,\eta \bm x)
            ]\\
        &=  - \eta_A\bm{\nabla}_{\eta x} a_0 (t,\bm x) 
            + \eta_A\eta \p_{-t} \bm{a} (t,\bm x)\\
        &\qquad\qquad
            - \im\eta_A\beta\bm\nabla_{\eta x} \p_t A_0 (t,\bm x)
            + \im\eta_A\eta \beta\p_{-t} \p_t \bm{A} (t,\bm x)\\
        &=  \eta_A\eta \qty[
                \bm{e}(t,\bm x)
                + \im\beta\p_t\bm{E}(t,\bm x)
            ],
    \end{aligned}\\
    &\begin{aligned}
        \mathrm{T}_\Theta\cdot\mathrm{KMS}_\Theta
            [\tilde{\bm{b}}(t,\bm x)]
        &=  \mathrm{T}_\Theta
                [\bm{\nabla}_x\times\tilde{\bm{a}}(t,\bm x)]\\
        &=  \mathrm{T}_\Theta\{
                -\bm{\nabla}_x\times\eta_A\eta[\bm{a}(-t,\eta \bm x)+\im\beta\p_{-t}A(-t,\eta \bm x)]
            \}\\
        &=  - \bm{\nabla}_{\eta x}\times\eta_A\eta
                [\bm{a}(t,\bm x) + \im\beta\p_tA(t,\bm x)]\\
        &=  - \eta_A \qty[
                    \bm{b}(t,\bm x)
                    + \im\beta \p_t\bm{B} (t,\bm x)
                ].
    \end{aligned}
\end{align}\end{subequations}
Based on these, one can compute the dynamical KMS transformation of various terms.
For example, the dynamical KMS transformation of the term $\bm e \cdot \bm E$ (eq.~\eqref{DKMS_eE} in the main text) is computed as 
\begin{equation}\begin{split}
    &\mathrm{T}_\Theta\cdot\mathrm{KMS}_\Theta
        [\bm{e}\cdot\bm{E}(t,\bm x)]
    - \bm{e}\cdot\bm{E}(t,\bm x)\\
    &= \mathrm{T}_\Theta [(\bm{e}(-t,\eta \bm x) + \im\beta\p_{-t}\bm{E}(-t,\eta \bm x)) \cdot\bm{E}(-t,\eta \bm x)]
        - \bm{e}\cdot\bm{E}(t,\bm x)\\
    &= (\bm{e}(t,\bm x) + \im\beta\p_t\bm{E}(t,\bm x)) \cdot\bm{E}(t,\bm x)
        - \bm{e}\cdot\bm{E}(t,\bm x)\\
    &= \p_t\qty(\frac{\im}{2}\beta\bm{E}^2(t,\bm x)),
\end{split}\end{equation}
where we have explicitly indicated the spacetime coordinate dependence of the fields.

\subsection{Construction of dynamical KMS symmetric terms}\label{app:DKMS-symmetric-terms}

In this subsection, we explain how to systematically construct operators that are invariant under the dynamical KMS symmetry, up to total derivatives.
An operator $\Ocal[a,A]$ is said to be \emph{dynamical KMS symmetric} if it satisfies
\begin{equation}
    \mathrm{T}_\Theta\cdot\mathrm{KMS}_\Theta
        \{\Ocal[a,A]\}
    - \Ocal[a,A]
    = \p_\mu V_{\Ocal}^\mu [a,A],
\end{equation}
for some function $V_{\Ocal}^\mu [a,A]$.
An operator $\Ocal[a,A]$ can be decomposed into $\Theta$-even and $\Theta$-odd parts,
\begin{equation}\begin{split}
    \Ocal_{\rm e}[a,A]
    &\coloneqq \frac{1}{2}\qty(
            \Ocal[a,A]
            + \mathrm{T}_\Theta\cdot\Theta\{\Ocal[a,A]\}
        ),\\
    \Ocal_{\rm o}[a,A]
    &\coloneqq \frac{1}{2}\qty(
            \Ocal[a,A]
            - \mathrm{T}_\Theta\cdot\Theta\{\Ocal[a,A]\}
        ).
\end{split}\end{equation}
A crucial property of the dynamical KMS transformation is that it generates terms of lower order in $a$-type fields, but never higher-order ones (see e.g., eqs.~\eqref{DKMS_eE} and \eqref{DKMS_ee}). 
In particular, the transformation of the $\Theta$-odd part takes the form
\begin{equation}
    \mathrm{T}_\Theta\cdot\mathrm{KMS}_\Theta
        \{\Ocal_{\rm o}[a,A]\}
    - \Ocal_{\rm o}[a,A]
    = - 2\Ocal_{\rm o}[a,A] + (\text{lower order in }a).
\end{equation}
This implies that $\Theta$-odd operators cannot appear as the highest-order terms in $a$ within a dynamical KMS-symmetric combination.
Consequently, the highest-order contribution in $a$ of any such operator must be $\Theta$-even.%
\footnote{
    We choose the coefficient of the highest-order term in $a$-type fields to be real; an overall imaginary factor can be reinstated at the end to satisfy the unitarity condition~\eqref{Unitarity2}.
}

\begin{table}[tb]
\centering
\begin{tabular}{l|cccc}\hline
    &
    $\mathcal{T}$ &
    $\mathcal{PT}$ &
    $\mathcal{CT}$ &
    $\mathcal{CPT}$ \\ \hline
    $\bm{e}\cdot\bm{E},\ \bm{b}\cdot\bm{B},\ \bm{b}\cdot(\bm\nabla\times\bm{B}),\ \bm{e}^2,\ \bm{b}^2$ & $+$ & $+$ & $+$ & $+$\\
    $\bm{a}\cdot\bm{B},\ \bm{a}\cdot\bm{b},\ \bm{e}\cdot\dot{\bm{B}},\ \bm{b}\cdot\dot{\bm{E}}$ & $+$ & $+$ & $-$ & $-$\\
    $\bm{b}\cdot\bm{E}$ & $-$ & $-$ & $+$ & $+$\\
    $\bm{e}\cdot\dot{\bm{E}},\ \bm{b}\cdot\dot{\bm{B}}$ & $-$ & $-$ & $-$ & $-$\\
    $\bm{e}\cdot(\bm{E}\times\bm{B})$ & $-$ & $+$ & $+$ & $-$
    \\
    $\bm{b}\cdot(\bm{E}\times\bm{B})$ & $+$ & $+$ & $-$ & $-$\\\hline
\end{tabular}
\caption{Transformation properties of various terms for different choices of $\Theta$. A `$+$' (`$-$') indicates that the term in the left column is (is not) flipped in sign under the transformation specified in the top row.}
\label{tab:ThetaOddEven}
\end{table}

Let us first consider operators whose highest order in $a$ is $a^1$.
Terms of order $a^0$ are forbidden by the unitarity condition~\eqref{Unitarity1}, so the $a^1$ operators listed in table~\ref{tab:ThetaOddEven} must be combined so as to satisfy the dynamical KMS condition.
This analysis yields the following allowed combinations,
\begin{equation}
    \bm{e}\cdot\bm{E},\
    \bm{b}\cdot\bm{B},\
    \bm{a}\cdot\bm{B},\
    (\bm{e}\cdot\dot{\bm{B}}
     + \bm{b}\cdot\dot{\bm{E}}).
    \label{DKMSsym_a1}
\end{equation}
These operators are dynamical KMS symmetric provided that they are $\Theta$-even.
In the present case, each term is individually invariant, although in general a nontrivial linear combination may be required.
Other $a^1$ operators, even when they are $\Theta$-even, transform as
\begin{equation}\begin{split}
    \mathrm{T}_\Theta\cdot\mathrm{KMS}_\Theta
        [\bm{e}\cdot\bm{B}]
    - \bm{e}\cdot\bm{B}
    &= \im\beta\dot{\bm{E}}\cdot\bm{B},\\
    \mathrm{T}_\Theta\cdot\mathrm{KMS}_\Theta
        [\bm{e}\cdot(\bm{E}\times\bm{B})]
    - \bm{e}\cdot(\bm{E}\times\bm{B})
    &= \im\beta\dot{\bm{E}}\cdot(\bm{E}\times\bm{B}),\\
    \mathrm{T}_\Theta\cdot\mathrm{KMS}_\Theta
        [\bm{b}\cdot(\bm{E}\times\bm{B})]
    - \bm{b}\cdot(\bm{E}\times\bm{B})
    &= \im\beta\dot{\bm{B}}\cdot(\bm{E}\times\bm{B}),
\end{split}\end{equation}
Since these variations cannot be written as total derivatives, even after taking linear combinations, such terms are excluded in the effective theory.

We now turn to operators whose highest order in $a$ is $a^2$.
In this case, $\Theta$-even $a^2$ terms can combine with $\Theta$-odd $a^1$ terms to form dynamical KMS-symmetric structures.
This feature originates from the imaginary shift in the dynamical KMS transformation, schematically $a \mapsto a + \im\beta \dot{A}$.
The allowed combinations are
\begin{equation}
    \bm{e}\cdot(\bm{e}+\im\beta\dot{\bm{E}}),\
    \bm{b}\cdot(\bm{b}+\im\beta\dot{\bm{B}}),\
    \bm{b}\cdot(\bm{a}+\im\beta\bm{E}).
    \label{DKMSsym_a2}
\end{equation}
These exhaust all possible independent structures, since each operator can appear at most once as the highest-order term.
Although linear combinations of $a^2$ terms are in principle allowed, this is not the case here.

So far we have assumed that the $\Theta$ symmetry is unbroken.
However, one can consider a situation where the $\Theta$ symmetry is spontaneously broken (see the next subsection for more details).
Since $\Theta$ is a $\mathbb{Z}_2$ symmetry,
the theory admits two degenerate vacua separated by an energy barrier.
At low energies, the dynamics is described by effective field theories around each vacuum, characterized by the order parameter $c$ introduced in eq.~\eqref{DKMS_transf_SSB}.
Under $\Theta$ and $\mathrm{KMS}_\Theta$, this order parameter transforms as
\begin{equation}
    \Theta c = -c, \qquad
    \mathrm{KMS}_\Theta[c] = -c.
\end{equation}
Multiplication by $c$ therefore converts $\Theta$-odd operators into $\Theta$-even ones.
As a result, additional contributions to the effective Lagrangian become allowed, built from the operators listed in
eqs.~\eqref{DKMSsym_a1} and \eqref{DKMSsym_a2}, within $m+n\leq4$.

\subsection{Spontaneously broken \texorpdfstring{$\Theta$ symmetry}{Theta}}
\label{app:DKMS_SSB}

Here, we discuss how the dynamical KMS transformation is formulated when a discrete symmetry $\Theta$ involved in it is spontaneously broken.

Whether it is continuous or discrete, when the symmetry is spontaneously broken, we need to be careful about the order of two limits:
\begin{equation}
    \rho \coloneqq
    \frac{1}{Z_0}
    \lim_{\delta\to+0}\qty(
        \lim_{V\to\infty}
        e^{-\beta H_V^{(\delta)}}
    ),\quad
    H_V^{(\delta)} = \int_V \mathcal{H} + \delta \mathcal{V},
\end{equation}
where $V\to\infty$ is thermodynamic limit and $\delta\to +0$ is the vanishing limit of the explicit breaking term $\delta \mathcal{V}$.
Here, the Hamiltonian $H_V^{(0)}=\int_V \mathcal{H}$ possesses the symmetry.
When the density matrix defined in this manner is not invariant under the symmetry transformation, the spontaneous symmetry breaking occurs.

Both in the in-out (S-matrix) and in-in (Schwinger-Keldysh) formalisms, the real-time evolution is considered.
It is expected that the procedure $\lim_{V\to\infty}e^{\pm itH_V^{(\delta)}}$ does not induce any singularity at $\delta\sim 0$ so that the unitary evolution is invariant under the symmetry transformation for $\delta=0$.
This is the reason why we impose the full symmetry for the effective action even when the symmetry is spontaneously broken.

If the spontaneously broken symmetry is continuous ($\mathcal G\to\mathcal H$), the low-energy excitations are the Nambu–Goldstone modes. 
The standard construction of effective field theory relies on the parametrization of the coset space, $\mathcal G/\mathcal H$. 
Therefore, after taking $\delta\to +0$, the information carried by $\delta\neq 0$ is absorbed into the nonlinear realization of the symmetry $\mathcal G$.
If the spontaneously broken symmetry is discrete, gapless excitations are not guaranteed to exist. 
In this case, however, even after taking $\delta\to +0$, the order parameter persists as an imprint of the microscopic $\delta\neq 0$
theory and behaves as a ``charged'' constant under the discrete symmetry in the low-energy sector. 
Therefore, if both continuous and discrete symmetries are spontaneously broken, and if the Nambu–Goldstone modes carry ``charges'' (i.e., transform nontrivially) under the broken discrete symmetry, then the order parameter can serve as an external field in addition to the other external parameters.

By this observation, we can generalize the dynamical KMS transformation when the discrete $\Theta$ symmetry is spontaneously broken.
In its derivation, we need to evaluate
\begin{equation}
    \rho_\Theta
    \coloneqq \Theta\rho\Theta^{-1}
    =
    \frac{1}{Z_0}
    \lim_{\delta\to+0}\qty[
        \lim_{V\to\infty}
        e^{-\beta\int_V(\mathcal{H} + \Theta\delta V\Theta^{-1})}
    ].
\end{equation}
We add an explicit breaking term of $\Theta$ symmetry, which is $Z_2$, so that $\Theta\delta \mathcal{V}\Theta^{-1} = -\delta \mathcal{V}$.
Therefore, we have 
\begin{align}
    \rho_\Theta
    =
    \frac{1}{Z_0}
    \lim_{\delta\to -0}\qty[
        \lim_{V\to\infty}
        e^{-\beta H_V^{(\delta)}}
    ].
\end{align}
We have an order parameter $c$ of $\Theta$ symmetry breaking, which is $\mathbb{Z}_2$ under $\Theta$.
When the order parameter is $c$ in $\delta\to +0$, it flips the sign to $-c$ in $\delta\to -0$.
The dynamical KMS transformation eqs.~\eqref{DKMS_transf} and \eqref{DKMS_Leff} is generalized as
\begin{subequations}
\begin{align}
    & W[\phi_+,\phi_-;c] = W[\tilde\phi_+,\tilde\phi_-;-c],\\
    & \Lcal_{\rm eff}[\tilde\phi_+,\tilde\phi_-,\tilde\chi_+,\tilde\chi_-;-c]
    - \Lcal_{\rm eff}[\phi_+,\phi_-,\chi_+,\chi_-;c]
    = \p_\mu V^\mu[\phi_+,\phi_-,\chi_+,\chi_-],
\end{align}
\label{DKMS_transf_SSB}
\end{subequations}
where the transformation of the fields is defined in eq.~\eqref{DKMS_transf_fields} for $W$ and eq.~\eqref{DKMStrsf0form} for $\Lcal_{\rm eff}$.

\bibliographystyle{JHEP}
\bibliography{References.bib}

\end{document}